\documentclass[pre,showkeys,preprintnumbers,amsmath,amssymb,amsfonts,superscriptaddress,twocolumn]{revtex4}

\usepackage{graphicx}
\usepackage{color}
\usepackage{bm}
\usepackage{ulem}

\def\ve{\varepsilon}

\def\R{{\mathbb R}}
\def\x{{\mathbf{x}}}
\def\A{{\mathcal A}}
\def\Rve{{\mathcal R}}

\def\erfc{\mathrm{erfc}}
\def\erfcx{\mathrm{erfcx}}

\begin{document}

\title{Towards a full quantitative description of single-molecule reaction
kinetics in biological cells}

\author{Denis S. Grebenkov}
  \email{denis.grebenkov@polytechnique.edu}
\affiliation{Laboratoire de Physique de la Mati\`{e}re Condens\'{e}e (UMR 7643),
CNRS -- Ecole Polytechnique, University Paris-Saclay, 91128 Palaiseau, France}
% ORCID 0000-0002-6273-9164

\author{Ralf Metzler}
  \email{rmetzler@uni-potsdam.de}
\affiliation{Institute of Physics \& Astronomy, University of Potsdam,
14476 Potsdam-Golm, Germany}
% ORCID 0000-0002-6013-7020

\author{Gleb Oshanin}
  \email{oshanin@lptmc.jussieu.fr}
\affiliation{ Sorbonne Universit\'e, CNRS, Laboratoire de Physique Th\'eorique de la Mati\`ere
Condens\'ee (UMR CNRS 7600), 4 Place Jussieu, 
F-75005, Paris, France}
% ORCID 0000-0001-8467-3226

\begin{abstract}
The first-passage time (FPT), i.e., the moment when a stochastic
process reaches a given threshold value for the first time, is a
fundamental mathematical concept with immediate applications.  In
particular, it quantifies the statistics of instances when
biomolecules in a biological cell reach their specific binding sites
and trigger cellular regulation. Typically, the first-passage
properties are given in terms of mean first-passage times. However,
modern experiments now monitor single-molecular binding-processes in
living cells and thus provide access to the full statistics of the
underlying first-passage events, in particular, inherent cell-to-cell
fluctuations. We here present a robust explicit approach for obtaining
the distribution of FPTs to a small partially-reactive target 
in cylindrical-annulus domains, which represent typical bacterial and
neuronal cell shapes. We investigate various asymptotic behaviours of
this FPT distribution and show that it typically is very broad in many
biological situations: thus, the mean FPT can differ from the most
probable FPT by orders of magnitude. The most probable FPT is shown to
strongly depend only on the starting position within the geometry and
to be almost independent of the target size and reactivity. These
findings demonstrate the dramatic relevance of knowing the full
distribution of FPTs and thus open new perspectives for a more
reliable description of many intracellular processes initiated by the
arrival of one or few biomolecules to a small, spatially localised
region inside the cell.
\end{abstract}


\date{\today}
%\footnote{This paper should be cited as: D. S. Grebenkov, R. Metzler, and G. Oshanin, Phys. Chem. Chem. Phys. {\bf 20}, 16393-16401 (2018).}

\maketitle

\section{Introduction}

Many intracellular processes of signalling, regulation, infection,
immune reactions, metabolism, or transmitter release in neurons are
triggered by the arrival of one or few biomolecules to a small
spatially localised region \cite{alberts,snustad}. Such processes
determine the cellular function and are controlled by the statistics
of the first-passage time (FPT) to a reaction event (also called the
reaction time), i.e., the instant in time when the respective
molecules hit their target site for the first time and initiate
biochemical responses \cite{Redner,Metzler,Holcman13,Bressloff13,
Benichou11,Benichou14}. With modern techniques such as superresolution
microscopy, it is possible to monitor individual, single-molecular
biochemical regulation and production processes in living cells,
revealing, for instance, pronounced fluctuations of production events
of individual messenger RNA or proteins within a single cell as well
as striking differences of production patterns between genetically
identical cells \cite{xie,xie1,avo}.

Most available analytical results to quantify the first-passage
dynamics were obtained for the {\it mean} first-passage time (MFPT)
\cite{Holcman14,Ward93,sasha1,
Singer06a,Singer06b,Singer06c,Condamin07,Benichou08,Pillay10,Cheviakov10,osh2010,
Cheviakov12,Caginalp12,sasha2,sasha3,Grebenkov16,Guerin16,Grebenkov17,tal},
corresponding to the inverse of the mean rate constant conventionally
used in biochemistry. For a bounded domain the MFPT is typically
proportional to the domain volume, and it diverges as the target
region shrinks.  In particular, for the so-called narrow escape
problem, which pertains to a variety of situations when a diffusive
particle has to leave a bounded domain through a very small window on
its boundary \cite{Holcman14,Schuss07}, the MFPT determines the
characteristic decay time of the exponential right tail of the
distribution of the FPT, likewise, for the case of a small target
inside bounded circular domains \cite{Godec16,Godec16a}. This
signifies that the MFPT is dominated by rather rare, anomalously long
searching trajectories, and thus can be non-representative of the
actual behaviour, or, at least be not the only important
characteristic time-scale. Indeed, if a particle with diffusion
coefficient $D$ is released within a short distance $\delta$ to the
target, the relevant time scale would be $\delta^2/D$, whereas the
MFPT would be of the order of $L^2/D$, where $L$ is the size of the
domain.  As a consequence, in this case the kinetics of the
aforementioned biological processes will most likely be determined by
the most probable FPT, which can be orders of magnitude smaller than
the MFPT, a scenario recently called the few-encounter limit
\cite{Godec16}.  Moreover, it was shown that two FPT events in the
same system may be dramatically disparate
\cite{gleb,gleb1,gleb2}. In these common and biologically relevant
situations, the whole FPT distribution is needed to adequately
quantify the molecular process and to meaningfully extract the kinetic
parameters from measurements.

However, the exact FPT distribution is known only for few elementary
cases such as the FPT to a perfectly reactive target placed at the
centre of a spherical region or to its boundary, starting from a fixed
location \cite{Redner}. At the same time, already finding the
distribution of FPTs to a small target region on the otherwise
reflecting boundary of a sphere remains an open problem. To our
knowledge, the only nontrivial case, for which an exact FPT
distribution was recently derived, is that of an arc-shaped target on
the boundary of a disk \cite{Rupprecht15}. To study the FPT in more
complicated realistic geometries, some approximate techniques have
been developed, such as the uniform approximation \cite{Isaacson13}
and the asymptotically exact Newton-series approach \cite{Godec16}.
Otherwise, one resorts to the numerical analysis of the full FPT
distribution \cite{Haffner2018}.  We emphasise that an impact of a
finite reactivity on the form of the FPT distribution remains a
completely open question.

\begin{figure}
\begin{center}
\includegraphics[width=60mm]{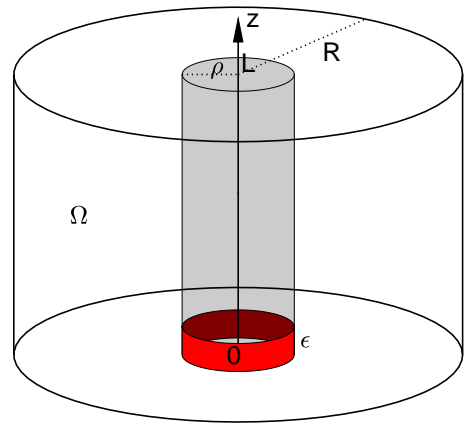}
\end{center}
\caption{Schematic presentation of the cylindrical-annulus domain $\Omega$
between two concentric cylinders of radii $\rho$ and $R$ and capped by
planes at $z=0$ and $z=L$. The target region is the red annulus of
radius $\rho$ and height $\epsilon$.}
\label{fig:scheme}
\end{figure}

We here report the approximate, but explicit and very accurate
expression for the distribution of the FPT to a partially-reactive
annular target on a cylinder, surrounded by a larger impermeable
cylinder and capped by two parallel planes (Fig. \ref{fig:scheme}),
which is the relevant geometry to describe the first passage of
molecules to the nucleoid region of bacteria cells or to a central
filament trail in the axon of a neuronal cell.  Another example of
such a geometry is provided by a usual experimental set-up for the
analysis of a diffusive search by a transcription factor protein for a
specific binding site on a single strand of elongated DNA (inner
cylinder), with the outer cylinder being the wall of the container.
We also note that from a mathematical point of view, the method
underlying the derivation of this FPT distribution can be formally
generalised to arbitrary bounded domains with a small target region,
and thus become applicable to the narrow escape problem in presence of
a barrier at the escape window.

Our solution relies on the self-consistent approximation (SCA)
technique originally devised by Shoup, Lipari, and Szabo
\cite{Shoup81} for the analysis of reaction rates between particles
with inhomogeneous reactivity, and recently applied to the MFPT in
spherical \cite{Grebenkov17} and cylindrical geometries
\cite{Grebenkov17a}.  Within this approximation, the exact mixed
boundary condition is replaced by an effective one, reducing the
problem to finding self-consistent solutions.  We adapt this
approximation to the modified Helmholtz equation governing the
survival probability in Laplace domain and thus the FPT distribution,
which is subsequently checked against the numerical solution of the
original problem, and is shown to be in a remarkable agreement with
the latter.  We note that the symmetries of the geometry under study
permit us to express the FPT distribution in a compact form under
rather general conditions: for arbitrary radii of the inner and the
outer cylinders, for arbitrary starting points, fixed or averaged over
the volume or over the cylindrical surface of a given radius, and for
an arbitrary chemical reactivity $\kappa$ defining the probability of
a reaction with the target upon encounter.

\begin{figure}
\begin{center}
\includegraphics[width=92mm]{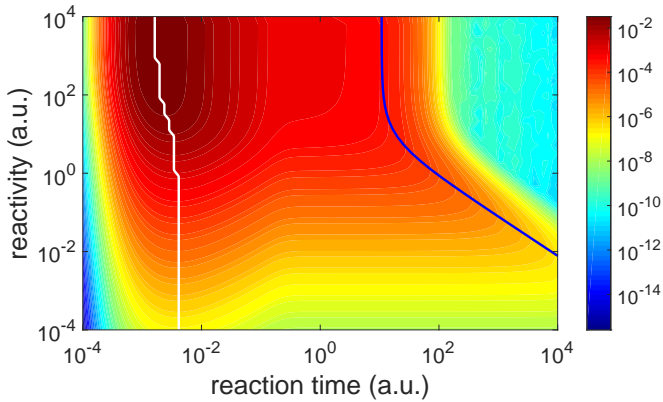}
\end{center}
\caption{
Impact of the finite reactivity onto the FPT probability density
(shown as a ``heatmap'', in which the value of the FPT density is
determined by the colour code). When the reactivity decreases, the
distribution becomes much broader and extends toward longer reaction
times.  Blue and white curves show respectively the mean and the most
probable FPTs versus the reactivity, and differ by orders of
magnitude.  The FPT probability density was obtained via a
numerical Laplace inversion of the solution \eqref{eq:H_tilde}.}
\label{fig:heatmap}
\end{figure}

We illustrate various features of this distribution, e.g., its
progressive broadening as the outer cylinder is becoming larger, or
the size of the target region is getting smaller, and highlight the
relevance of the most probable FPT. In addition, our analysis unveils
remarkable effects of the chemical reactivity $\kappa$ on the
functional shape of the FPT distribution which were not studied
systematically before (Fig. \ref{fig:heatmap}). In particular, we
proceed to show that upon lowering $\kappa$, a plateau-like region
develops beyond the most probable FPT, such that, interestingly, the
values of the FPT in an interval ranging over several decades turn out
to be almost equally probable (see Fig.  \ref{kappa}).  Moreover, the
chosen shape of a capped cylindrical annulus allows us to explore
various features of {\it effectively} one- (semi-infinite cylindrical
annulus), two- (exterior of a capped cylinder), and three-dimensional
(exterior of a semi-infinite cylinder) search in unbounded domains,
for which the MFPT is infinite.  In particular, we recover the
characteristic right tails $t^{-3/2}$ and $1/(t\ln^2 t)$ of the FPT
distribution in effectively one- and two-dimensional geometries
\cite{Redner}. Therefore, our analysis also provides a seminal
unifying framework in which the behaviour specific to one-, two-
and three-dimensional unbounded systems appears in particular
limits. Overall, our results emphasise an absolute necessity of
studying the first passage phenomena in biologically relevant systems
beyond the MFPT and mean rates, and show that the knowledge of the
full FPT distribution is indeed indispensable for getting a complete
understanding of the wealth of kinetic behaviour in such systems.

\section{Results}

We study the distribution of the FPT to an annular reactive region
$\Gamma$ (the target site) on a cylinder of radius $\rho$ when
diffusion is restricted by an outer, concentric and impermeable
cylinder of radius $R$ (Fig.~\ref{fig:scheme}).  In other words,
diffusion occurs within the confining cylindrical-annulus domain
$\Omega$ spanned by the interval $z\in(0,L)$ along the cylinder axis
and the radius $r=\sqrt{x^2+y^2}$ in the interval $r\in(\rho,R)$. The
target consists of the annular domain $\Gamma$ on the inner cylinder,
specified by the interval $z\in(0,\epsilon)$ along the cylinder axis
and the radius $r=\rho$. Note that the cylindrical domain is capped by
reflecting planes at $z=0$ and $z=L$ so that the scenario is in fact
equivalent to diffusion in an infinite cylinder with a periodic
arrangement of targets.  For a particle seeded at some point
$\x\in\Omega$, the survival probability $S(\x,t)=\int_{\Omega}
P(\x,\x',t)d\x'$ is calculated as the volume integral of the
(non-normalised) density function $P(\x,\x',t)$ to find the particle
at position $\x'$ at time $t$. The derivative of the survival
probability with respect to time, taken with sign minus, then produces
the probability density of first passage time,
$H(\x,t)=-\partial S(\x,t)/\partial t$.  In Laplace domain, defined in
terms of $\tilde{f}(p)=\int_0^{\infty}\exp(-pt)f(t)dt$, this relation
can be rewritten as $\tilde{H}(\x,p)=1-p\tilde{S}(\x,p)$, where we
used the initial condition $S(\x,t=0)=1$, that is, initially the
particle is present in the domain $\Omega$ with unit
probability.

The survival probability, written in cylindrical coordinates
$(r,z,\varphi)$, satisfies the backward Fokker-Planck equation
$\partial S(\x,t)/\partial t=D\Delta S$, where $D$ is the bulk
diffusion coefficient, and $\Delta=\partial_r^2+r^{-1}
\partial_r+\partial^2_z+r^{-2}\partial^2_\varphi$ is the Laplace
operator.  In Laplace domain, this equation reduces to the modified
Helmholtz equation 
\begin{equation} 
(p-D\Delta)\tilde{S}(\x,p)=1.  
\end{equation}
Due to the axial symmetry of the problem, there is no dependence on
the polar angle $\varphi$. The reflecting boundary conditions at the
outer boundaries are taken into account by setting the derivatives
$\partial S/\partial r=0$ at $r=R$ and $\partial S/\partial z=0$ at
$z=0,L$, respectively. To simplify notations, we replace the axial
coordinate $z$ by $\theta=\pi z/L$, and introduce
$\ve=\pi\epsilon/L$. The mixed boundary condition on the inner
cylinder then reads
\begin{equation}
\label{eq:BC_original}
D(\partial\tilde{S}/\partial r)_{r=\rho}=\begin{cases}\kappa
\tilde S_{r =\rho}\quad(0<\theta<\ve)\cr 0  \hskip 11mm (\ve<\theta<\pi)\end{cases} 
\end{equation}
in Laplace domain. The reactivity coefficient $\kappa$ in the Robin
boundary condition determines the degree of stickiness of the reactive
boundary $\Gamma$ and is associated with the probability of the
reaction with the target upon an encounter
\cite{Grebenkov03,Grebenkov06,Grebenkov07}.  In standard terms,
$\kappa$ (in units m/s) is defined as the rate describing the
number of reaction acts per unit of time within the volume of the
reaction zone around $\Gamma$, times the reaction radius and hence, is
a material property independent of $\epsilon$ (see \cite{diet} for
more details).  For a non-reactive target one has $\kappa=0$, while
$\kappa=\infty$ corresponds to the case of a perfect reaction which,
on encounter, occurs with probability $1$.  We note that the effect of
$\kappa$ on the shape of the full distribution of the FPT is a novel
feature here.  The only available previous analysis concerned solely
its effect on the MFPT, and showed that in related settings it can
indeed be decisive
\cite{Grebenkov17,Grebenkov17a}.  This naturally raises the question
of the effects of a finite reactivity beyond the MFPT.

We apply a SCA by replacing the mixed boundary condition
\eqref{eq:BC_original} by the inhomogeneous Neumann condition
\cite{Shoup81}
\begin{equation}
\label{eq:BC_modified}
D(\partial\tilde S/\partial r)_{r=\rho}=Q\Theta(\ve-\theta),
\end{equation}
in which $\Theta(z)$ is the Heaviside step function and the effective
flux $Q$ remains to be determined by imposing an appropriate
self-consistent closure relation \cite{Shoup81}, i.e., by requiring
that the first line in \eqref{eq:BC_original} holds on average, i.e.,
$D\int^{\epsilon}_0d\theta(\partial\tilde S/\partial r)_{r=\rho}=\kappa
\int^{\epsilon}_0 d\theta\tilde S_{r=\rho}$.

We search a solution in the generic form
\begin{equation}
\label{eq:ansatz}
\tilde{S}(r,\theta;p)=\frac{R^2}{D}\biggl(u_0(r)+\sum\limits_{n=0}^\infty a_n
g_n(r)\cos n\theta\biggr),
\end{equation}
where the first term is the solution of the inhomogeneous problem with Dirichlet
boundary condition at $r=\rho$, $a_n$ are unknown coefficients to be determined,
and $g_n(r)$ are radial functions satisfying the ordinary differential equation
\begin{equation}
\label{eq:gn_eq}
g''_n+\frac{1}{r}g'_n-\biggl(\frac{\pi^2n^2}{L^2}+\frac{s}{R^2}\biggr)g_n=0,
\end{equation}
where the prime denotes radial derivative and $s=pR^2/D$ is the dimensionless
Laplace variable. We emphasise the dependence of $u_0(r)$, $a_n$ and $g_n(r)$
on the Laplace variable, although we do not write it explicitly for the sake
of brevity.

The solution of \eqref{eq:gn_eq} satisfying the boundary condition $(\partial
g_n/\partial r)_{r=R} = 0$ is a linear combination of modified Bessel functions
$I_n(z)$ and $K_n(z)$ of first and second kind,
\begin{equation}
\label{eq:gn_free}
g_n(r)=I_0(\alpha_nr/L)K_1(\alpha_nR/L)+K_0(\alpha_nr/L)I_1(\alpha_nR/L),
\end{equation}
with $\alpha_n=\sqrt{\pi^2n^2+sL^2/R^2}$. The solution of the
inhomogeneous problem with Dirichlet boundary condition at $r=\rho$
reads $u_0(r)=[1-g_0(r)/g_0 (\rho)]/s$ \cite{Redner,Grebenkov17b}. The
coefficients $a_n$ are determined in the Supplemental Information
(SI), section \ref{coeffs}, and we obtain the final result for the FPT
density
\begin{eqnarray}
\label{eq:H_tilde}
\tilde{H}(r,\theta;p)=\eta\frac{g_0(r)}{g_0(\rho)}+2\eta\frac{g'_0(\rho)}{g_0(
\rho)}\sum\limits_{n=1}^\infty\frac{g_n(r)}{g'_n(\rho)}\frac{\sin n\ve}{n\ve}
\cos(n\theta),
\end{eqnarray}
in Laplace domain, where 
\begin{equation}
\label{eq:eta}
\eta=\biggl(1-\biggl(\frac{\pi D}{\kappa\ve}+\frac{L}{\pi}\Rve_\ve\biggr)
\frac{g'_0(\rho)}{g_0(\rho)}\biggr)^{-1} 
\end{equation}
and
\begin{equation}
\label{eq:Rve}
\Rve_\ve=-\frac{2\pi}{L}\sum\limits_{n=1}^\infty\frac{g_n(\rho)}{g'_n(\rho)}
\biggl(\frac{\sin n\ve}{n\ve}\biggr)^2 .
\end{equation}

This approximate representation of the FPT density in the
cylindrical-annulus domain is one of the main results of this
paper. This result hinges on the SCA, which has already been applied
for the analysis of the MFPT in spherical \cite{Grebenkov17} and
cylindrical-annulus \cite{Grebenkov17a} geometries, and verified
against the numerical solution of the mixed boundary
problem. Moreover, a similar SCA approach has been used in \cite{diet}
to calculate the self-propulsion velocity of catalytically-active
colloids and was shown to be in very good agreement with already known
results, only slightly underestimating some insignificant numerical
factors.  In Sec. \ref{sec:validation} of the SI we show that it is a
remarkably accurate approximation for the problem under study,
checking it for different initial conditions against the numerical
solution of the original mixed boundary value problem.

The moments of this FPT distribution can be obtained from
$\tilde{H}(r,\theta;p)$ in the form ${T_n} = (-1)^n\lim_{p\to
0}(\partial^n/\partial p^n) \tilde{H}(r,\theta;p)$. The first
moment $T_1$ is the mean FPT that we also denote as $T$ for brevity.
The explicit solution in \eqref{eq:H_tilde} fully determines the
statistics of the FPT. Since $\tilde{S}(r,\theta;p)$ and the FPT
density $\tilde{H}(r,\theta;p)$ are trivially related in Laplace
domain, we focus on the latter quantity, bearing in mind that all
properties of the Laplace-transformed survival probability follow
immediately from those of $\tilde H(r,\theta;p)$. The inverse Laplace
transform can be performed either by determining the poles of
$\tilde{H}(r,\theta;p)$ and using the residue theorem, or by numerical
inversion using the Talbot algorithm. In Sec. \ref{sec:time} of the
SI, we discuss in more detail the former approach, whereas the
numerical inversion is used throughout the paper.  The solutions in
the limiting cases $R\to\infty$ and $L\to\infty$ are presented in the
SI (Sections \ref{sec:Rinf} and \ref{sec:Linf}).

As already remarked, we will consider different situations with respect to
the starting point of the particle.  If the starting point is distributed
uniformly in the bulk, the volume average of $\tilde{H}(r,\theta;p)$ can be
evaluated exactly,
\begin{equation}
\label{lap_ex}
\overline{\tilde H(p)}=2\int_0^{\pi}d\theta\int_{\rho}^R\frac{drr\tilde{H}(r,
\theta;p)}{\pi(R^2-\rho^2)}= \frac{- 2\rho g'_0(\rho)\eta}{s(1-\rho^2/R^2)g_0(\rho)},
\end{equation}
where we used the identity $\int_\rho^Rdrrg_n(r)=-\rho L^2g'_n(\rho)/\alpha_n^2$.
If in turn the average is taken over uniformly distributed starting points on a
cylindrical surface of radius $r$, we find
\begin{equation}
\label{eq:H_tilde_surf}
\overline{\tilde H(p)}_r=\frac{1}{\pi}\int\limits_0^\pi d\theta\tilde{H}(r,
\theta;p)=\eta \,\frac{g_0(r)}{g_0(\rho)}.
\end{equation}
Setting $r=\rho$ (when a particle starts from the inner boundary with
uniform density), this relation turns out to provide a natural
interpretation for the coefficient $\eta$ defined in
\eqref{eq:eta}.

As discussed in Ref.~\cite{Grebenkov17b}, $\tilde H(r,\theta;p)$ can
also be interpreted as the probability that a mortal walker with bulk
killing rate $p$ reaches the target. For $p=0$ the classical immortal
walker reaches the target with unit probability because of the
recurrent character of restricted Brownian motion in a bounded domain.
In turn, when $p>0$, the random walker can be killed during its search
for the target, and $\tilde H(r,\theta;p)$ is the fraction of walkers
that reach the target before being killed.

\section{Discussion}

The explicit form of the Laplace-transformed FPT distribution $\tilde{H}(r,
\theta ;p)$ in \eqref{eq:H_tilde} provides unprecedented opportunities for
studying the details of the first passage dynamics in a
cylindrical-annulus domain. The major challenge here is the relatively
large number of relevant parameters of this problem. In fact, the
short-time and the long-time behaviours of the FPT distribution (i.e.,
its left and right tails) strongly depend on the four geometric
parameters $R$, $L$, $\rho$, and $\epsilon$, as well as on the
reactivity $\kappa$, and on the starting point (in particular, whether
it is fixed or randomly distributed over some subspace). For instance,
the behaviour in the small-target limit $\epsilon\to0$ is expected to
be different from that in the thin cylinder limit $\rho\to0$.
Moreover, one can also investigate the limiting cases of the unbounded
exterior of a capped cylinder ($R\to\infty$), and of an infinitely
long cylinder ($L\to \infty$). In these two limits, the distribution
of the FPTs remains well defined, although the MFPT is infinite, as
shown in Sections~\ref{sec:Rinf} and \ref{sec:Linf} of the SI. We
discuss below the various facets of the FPT distribution in different
parameter ranges as well as some direct applications.

\subsection{General qualitative behaviour}
\label{sec:qualitative}

The form of the left tail of the FPT distribution (corresponding to
short FPTs) strongly depends on the starting point of the particle. If
the starting point is fixed (or surface-averaged with $r>\rho$), the
FPTs are dominated by very rare trajectories from $\x$ to the closest
points of the target (called direct trajectories in
Refs.~\cite{Godec16,Godec16a}). As $t\to 0$, we thus expect the
behaviour $H(\x,t)\propto\exp(-|\x-\Gamma|/[4Dt])$, where $|\x-\Gamma|$
is the Euclidean distance between the starting point $\x$ and
the target domain $\Gamma$.  In this limit, the FPT density vanishes
very rapidly, meaning that very short FPTs are extremely unlikely. In
turn, if the starting point is averaged over the volume or over the
inner surface at $r=\rho$, such that some particles are initially
released right at the surface of the target, one can expect that the
FPT density is peaked at $t = 0$ and then monotonically decreases with
$t$. In this case, an intermediate power law decay of the FPT
distribution is expected. In particular, the general asymptotic
behaviour derived in Ref.
\cite{Grebenkov17b} for the perfectly reactive target 
implies $\overline{\tilde{H}(p)} \simeq(|\Gamma|/|\Omega|)~(p/D)^{-1/2}$,
thus
\begin{equation}
\label{eq:Hvolume_t0}
\overline{H(t)}\simeq(2\rho\ve D)\pi^{-3/2}(R^2-\rho^2)^{-1}(Dt)^{-1/2}
\end{equation}
as $t\to0$. In the partially reactive case $\kappa<\infty$ the intermediate
power-law decay has a different form, see Sec.~\ref{sec:Limit_pinfty}.

The form of the right tail of the FPT distribution essentially depends on
whether the domain $\Omega$ is bounded or not. For any bounded domain, the
spectrum of the governing Laplace operator is discrete, and the FPT density
exhibits an exponential decay whose rate is determined by the smallest
non-trivial eigenvalue $\lambda_0$: $H(\x,t) \propto\exp(-Dt\lambda_0)$
as $t\to\infty$. In Sec.~\ref{sec:limitp0} we relate the decay rate to
the surface-averaged MFPT $\overline{T}_\rho$, which
is finite. The behaviour is different in the limits $R\to\infty$ or $L\to
\infty$ when the domain $\Omega$ becomes unbounded. In this case, the MFPT
is infinite, and the FPT density exhibits a power law decay (possibly with
logarithmic corrections). We discuss this behaviour in detail in the SI (see
Sections~\ref{sec:Rinf} and \ref{sec:Linf}). It is important to stress that
the related power law behaviour can also be relevant even for bounded domains
as an intermediate regime, before the ultimate exponential cut-off, see also
the findings for spherical domains in Ref. \cite{Godec16,Godec16a,gleb,gleb1,gleb2}.
As we will illustrate below, such an intermediate power law regime can spread
over a quite broad range of times and thus be the most interesting feature
of the underlying FPT phenomenon. In this situation, the most probable FPT
can differ from the MFPT by many orders of magnitude.

\subsection{The right-tail of the FPT distribution}
\label{sec:limitp0}

In the limit $p\to 0$ the Laplace transform $\tilde{H}(r,\theta;p)$ of
the FPT density determines both the moments $T_n$ of the FPT
and the long-time behaviour of $H(r,\theta;t)$ itself. Taking the
respective limits of the radial function discussed in
Sec.~\ref{long_short} of the SI, we obtain $\overline{H(t)}_{\rho}$ as
the inverse Laplace transformation of $\eta$, namely,
\begin{equation}
\overline{H(t)}_{\rho}\simeq\exp\bigl(-t/\overline{T}_\rho\bigr)/
\overline{T}_\rho,
\end{equation}
valid for $t\to\infty$. The characteristic time is given by
\begin{equation}
\label{surftau}
\overline{T}_\rho=\frac{R^2-\rho^2}{2D\rho}\biggl(\frac{\pi
D}{\kappa \ve} + \frac{L}{\pi} \Rve_\ve(p=0)\biggr),
\end{equation}
which corresponds to the surface-averaged MFPT investigated in
Ref. \cite{Grebenkov17a}.  This result is expected for diffusion in a
bounded domain. The asymptotic behaviour of other quantities can be
obtained in a similar way. For instance,
\begin{equation}
\overline{H(t)}_{r}\simeq\exp\bigl(-t/\overline{T}_r\bigr)/\overline{T}_r,
\end{equation}
with the characteristic time
\begin{equation}
\overline{T}_r=\overline{T}_\rho+\biggl(\frac{
\rho^2-r^2}{4D}+\frac{R^2 \ln(r/R)}{2D}\biggr),
\end{equation}
where the second term in the parentheses is the MFPT to the inner cylinder
from a uniformly distributed point at the cylindrical surface at $r$. The
additivity of two MFPTs reflects the fact that any trajectory from such a
point to the target can be split into two independent parts: the path from
the cylinder at $r$ to the cylinder at $\rho$, and the path from the cylinder
at $\rho$ to the target, similar to the results for inhomogeneous diffusion
in a cylindrical domain \cite{aljaz_add}.

\subsection{The left tail of the FPT distribution}
\label{sec:Limit_pinfty}

The form of the left tail of the FPT distribution stems from the
asymptotic behaviour of $\tilde H(r,\theta;p)$ in the limit
$p\to\infty$. After the transformations detailed in the SI (see
Section~\ref{long_short}), we obtain the Laplace-transformed FPT
density $\tilde H(r,\theta;p)$ along with its volume and surface
averages, $\overline{\tilde{H}(p)}$ and
$\overline{\tilde{H}(p)}_r$. Here one needs to distinguish the cases
of perfect ($\kappa=\infty$) and imperfect ($\kappa<\infty$)
reactivity at the target. Note that the difference in the asymptotic
behaviours for perfectly or only partially reactive targets was
discussed for other geometries in
Refs.~\cite{Grebenkov10a,Grebenkov10b}.

\subsubsection{Perfect reactions}

\begin{figure}
\includegraphics[width=92mm]{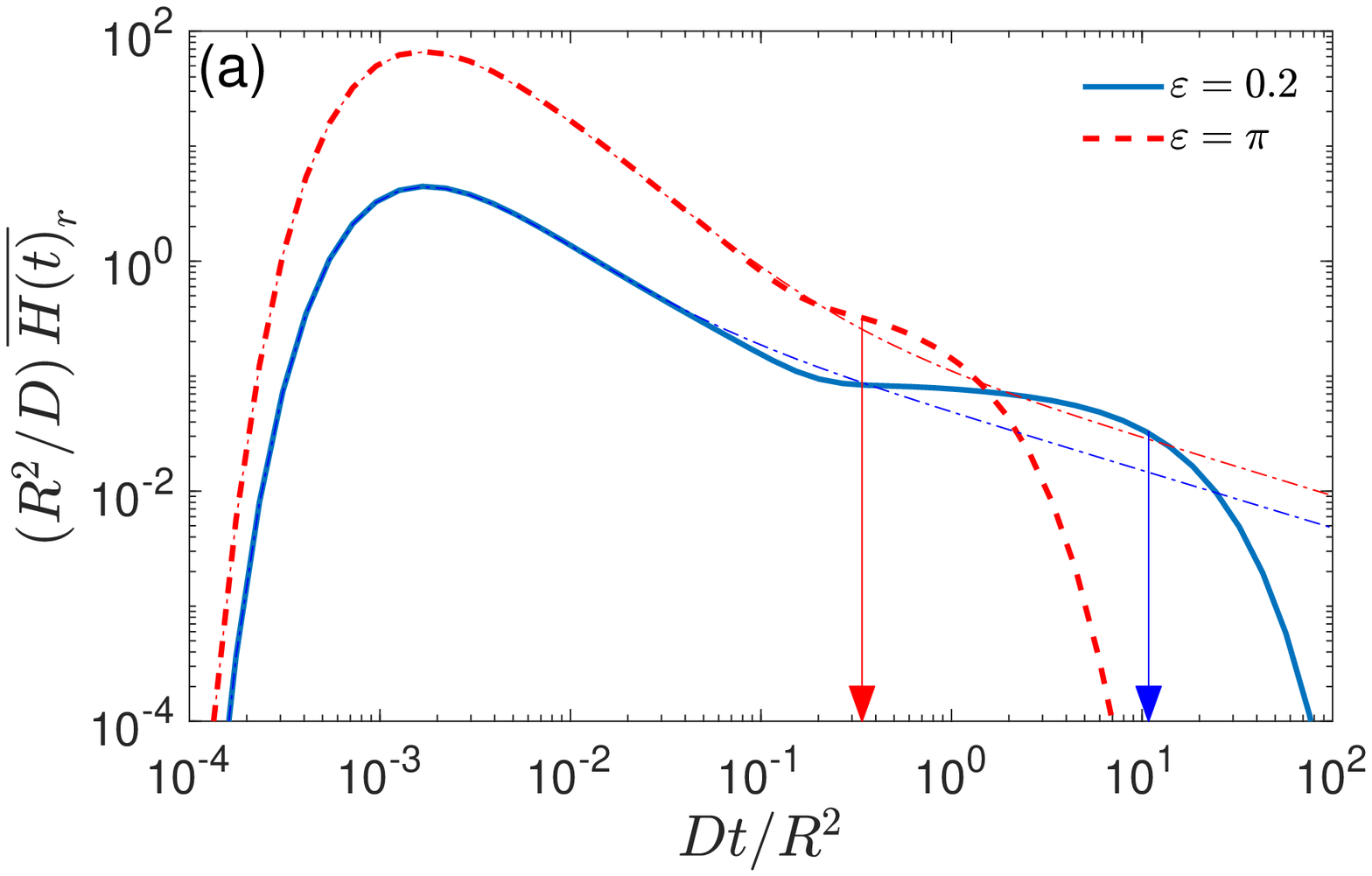}
\includegraphics[width=92mm]{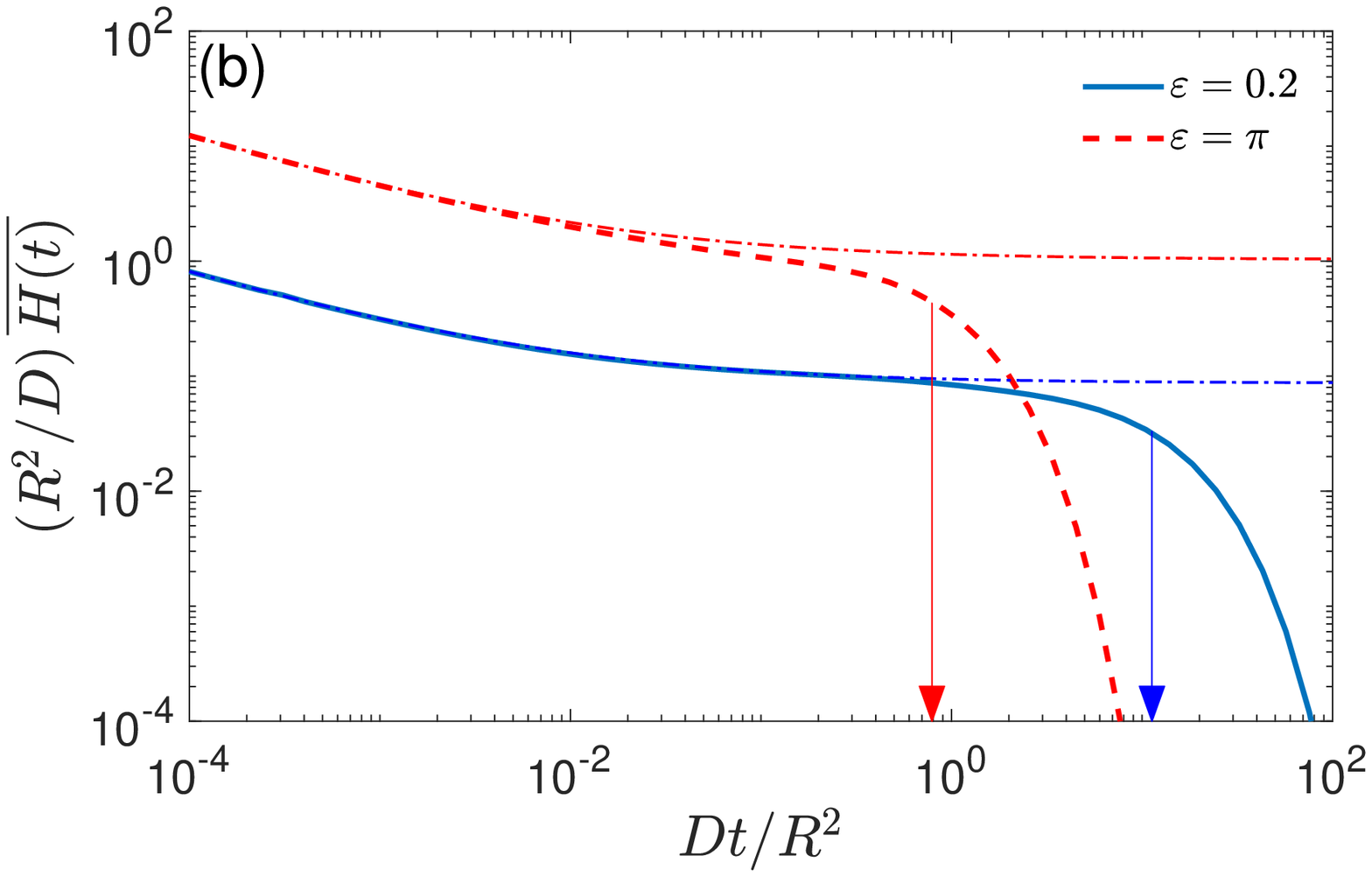}
\caption{
Surface-averaged {\bf (a)} and volume-averaged {\bf (b)} FPT densities
$\overline{H(t)}_r$ and $\overline{H(t)}$ as functions of $t$ for
perfect reactions ($\kappa=\infty$) with $L/R = \pi$, $\rho/R =0.1$,
$r/R =0.2$, and $\varepsilon=0.2$ (solid line) and $\varepsilon=\pi$
(dashed line).  Both curves are obtained by the numerical Laplace
inversion of \eqref{eq:H_tilde}. The two arrows indicate the MFPT $D
\overline{T}_r/R^2$ for both cases: $10.82$ ($\varepsilon=0.2$) and
$0.34$ ($\varepsilon=\pi$) for surface-averaged quantity, and $11.27$
($\varepsilon= 0.2$) and $0.79$ ($\varepsilon=\pi$) for
volume-averaged quantity.  Dash-dotted lines indicate the short-time
asymptotic \eqref{eq:Ht_r_t0} and \eqref{eq:Ht_t0_kappainf}, and agree
very well with the general result in \eqref{eq:H_tilde} well beyond
the most probable FPT.  Length and time scales are fixed by setting
$R=1$ and $R^2/D=1$.}
\label{fig:FPT3}
\end{figure}

According to \eqref{eq:H_tilde_surf} the inverse Laplace transform of
asymptotic \eqref{eq:eta_pinf} in the SI yields the asymptotic
behaviour of the surface-averaged FPT density $\overline{H(t)}_\rho$ at
small $t$, namely,
\begin{equation}
\overline{H(t)}_\rho\simeq(\ve/\pi)\delta(t)+(D/[8\pi])^{1/2}L^{-1}t^{-1/2}
+O(1).  
\end{equation}
The first term represents the fraction $\ve/\pi$ of particles, that
started right at the target, for which the first passage time is
zero. The next term accounts for the FPTs of particles with non-zero
initial separations from the target.  Since \eqref{eq:eta_large} was
derived for $\ve\leq\pi/2$, the above asymptotic behaviour is not
applicable for the case $\ve=\pi$, for which
$\overline{H(t)}_\rho=\delta( t)$ without correction terms.

When the particles start from a cylindrical surface at $r$, \eqref{eq:H_tilde_surf}
has an extra factor $g_0(r)/g_0(\rho)$. With the large-$p$ asymptotic
\eqref{eq:Hp_asympt} we find the short-time behaviour
\begin{eqnarray}
\nonumber
\hspace*{-1cm}
&&\overline{H(t)}_r\simeq\frac{\ve}{\pi}(\rho/[4\pi rDt^3])^{1/2}\exp(-(r-\rho)^2/
[4Dt])\\
\hspace*{-1cm}
&& \times \biggl(r-\rho+Dt\biggl(\frac{\pi}{\sqrt{2}L\ve}+\frac{\sqrt{
1/\rho}-\sqrt{1/r}}{4\sqrt{R}}\biggr)+O(t^2)\biggr).
\label{eq:Ht_r_t0}
\end{eqnarray}
Figure \ref{fig:FPT3}(a) shows the surface-averaged probability
density $\overline{H(t)}_r$ for two choices of the target
height: $\ve = 0.2$ and $\ve = \pi$.  The latter case describes the
whole inner cylinder as reactive, while the former value of $\ve$ is
chosen arbitrarily and meant to illustrate a moderately small target.
In both cases shown in the figure the short-time asymptotic
\eqref{eq:Ht_r_t0} is very accurate up to $Dt/R^2\lesssim0.1$.  When
the target is the entire inner cylinder ($\ve=\pi$) this time scale is
of the order of the corresponding MFPT
$D\overline{T}_r/R^2\approx0.34$.  For times of that order the FPT
density has an exponential cut-off. For the case of a partially
reactive inner cylinder ($\ve=0.2$) the MFPT is, notably, around four
decades longer than the most likely FPT.

For the volume average, \eqref{lap_ex}, together with
\ref{eq:eta_pinf}, yields a different short-time behaviour, namely,
\begin{equation}
\label{eq:Ht_t0_kappainf}
\overline{H(t)}\simeq\frac{\rho D}{R^2-\rho^2}\biggl(\frac{2\ve}{\pi\sqrt{\pi Dt}} 
+\biggl(\frac{1}{\sqrt{2}L}+\frac{\ve}{\pi\rho}\biggr)+O(t^{1/2})
\biggr).
\end{equation}
The leading term agrees with the general behaviour in
\eqref{eq:Hvolume_t0}.  Figure~\ref{fig:FPT3}(b) shows the FPT density
obtained by numerical inversion of $\overline{\tilde H(p)}$ from
\eqref{eq:H_tilde}. In the particular case $\ve =\pi$ (the entire
inner cylinder is absorbing), one has $a_n=0$ and thus
\eqref{eq:H_tilde} is exact. One can see that both distributions are
broad. The asymptotic \eqref{eq:Ht_t0_kappainf} is remarkably accurate
for both cases $\ve=\pi$ and $\ve = 0.2$.

\subsubsection{Imperfect reactions}

For imperfect reactions with finite reactivity $\kappa$ the
first arrival onto the target does not necessarily imply a successful
reaction, so that the reaction times are increased. Indeed, for
$\kappa<\infty$,
\eqref{eq:eta_large} in the SI acquires the asymptotic \eqref{eta_imp},
from which we get the short-time behaviours
\begin{subequations}
\label{imperfect}
\begin{eqnarray}
\overline{H(t)}_\rho&\simeq&\frac{\ve}{\pi}\frac{\kappa}{\sqrt{\pi Dt}}-\frac{
\ve}{\pi}\biggl(\frac{\kappa^2}{D}+\frac{\kappa}{2\rho}\biggr)+O\bigl(\sqrt{t}\bigr),\\
\label{eq:Ht_r_t0_kappa}
\overline{H(t)}_r&\simeq&\frac{\ve}{\pi}\sqrt{\rho/r}\frac{\kappa}{\sqrt{\pi Dt}}
\exp\biggl(-\frac{(r-\rho)^2}{4Dt}\biggr),\\
\label{eq:Ht_t0_kappa}
\overline{H(t)}&\simeq&\frac{2\rho\ve\kappa}{\pi(R^2-\rho^2)}\biggl(1-\frac{2
\kappa\sqrt{Dt}}{D\sqrt{\pi}}+O(t)\biggr).
\end{eqnarray}
\end{subequations}
Interestingly, for imperfect reactions the leading short-time behaviour
of the FPT distribution appears to be distinctly different, depending
on the starting point: $\overline{H(t)}_\rho$ diverges as $t \to 0$,
$\overline{H(t)}_r$ tends to zero in this limit, while
$\overline{H(t)}$ approaches a constant value.

\begin{figure}
\includegraphics[width=92mm]{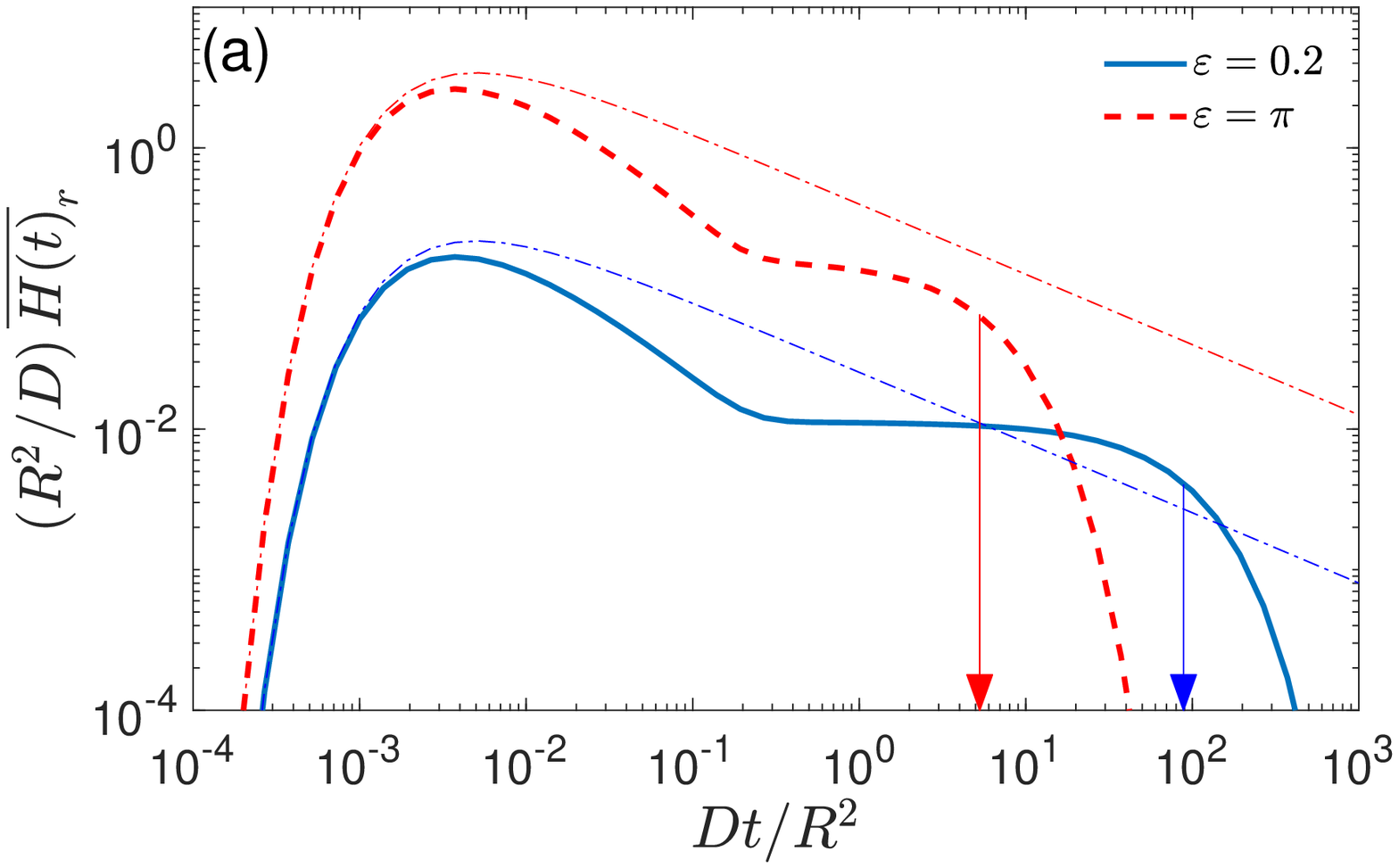}
\includegraphics[width=92mm]{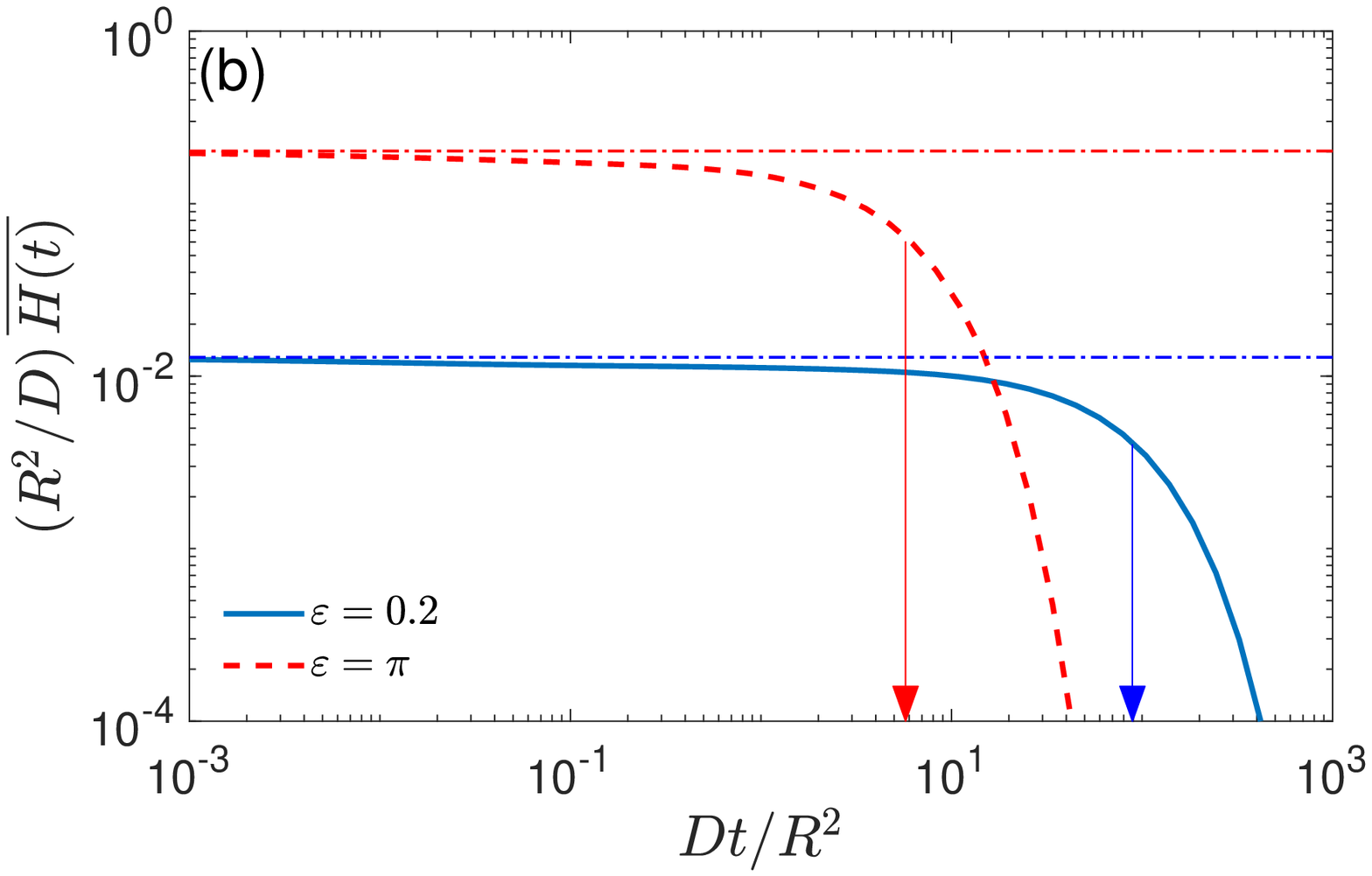}
\caption{
Surface-averaged {\bf (a)} and volume-averaged {\bf (b)} FPT densities
$\overline{H(t)}_r$ and $\overline{H(t)}$ as functions of $t$ for
imperfect reactions ($\kappa R/D = 1$, cf. Fig. \ref{fig:FPT3}), with
$L/R =\pi$, $\rho/R=0.1$, $r/R=0.2$, and $\varepsilon=0.2$ (solid
line) and $\varepsilon=\pi$ (dashed line). Both curves are obtained by
numerical Laplace inversion of \eqref{eq:H_tilde}. The two arrows
indicate the MFPT $D\overline{T}_r/R^2$ for both cases: $88.58$
($\varepsilon=0.2$) and $5.29$ ($\varepsilon=\pi$) for
surface-averaged quantity, and $89.03$ ($\varepsilon=0.2$) and $5.74$
($\varepsilon=\pi$) for volume-averaged quantity. Dash-dotted lines
show the short-time asymptotic \eqref{eq:Ht_r_t0_kappa} and
\eqref{eq:Ht_t0_kappa} (in which only the leading term is kept).
Length and time scales are fixed by setting $R = 1$ and $R^2/D = 1$.}
\label{fig:FPT3_kappa1}
\end{figure}

Figure~\ref{fig:FPT3_kappa1}(a) shows the surface-averaged FPT density
$\overline{ H(t)}_r$ at $r/R=0.2$ and $\kappa R/D=1$. One can see that
the short-time asymptotic \eqref{eq:Ht_r_t0_kappa} accurately
reproduces the behaviour of this density up to its maximum. As a
consequence, the position $t_m$ of this maximum can be obtained by
taking the derivative of
\eqref{eq:Ht_r_t0_kappa} with respect to $t$ and setting the resulting
expression equal to zero.  This gives the following estimate for the
most probable FPT:
\begin{equation}
\label{eq:tm}
t_m=(r-\rho)^2/(2D). 
\end{equation}
The estimated value of $t_m$ in \eqref{eq:tm} depends only on the
distance to the target but does not depend on either the target size
$\ve$ or the reactivity $\kappa$, nor on the inner radius of the
cylinder. In this example, $Dt_m/R^2=0.005$, whereas the MFPT
is four orders of magnitude higher. Similar to the findings in
Ref.~\cite{Godec16} the most likely FPT corresponds to
geometry-controlled, direct trajectories, in which the initial
distance from the target is decisive.

\begin{figure}
\includegraphics[width=92mm]{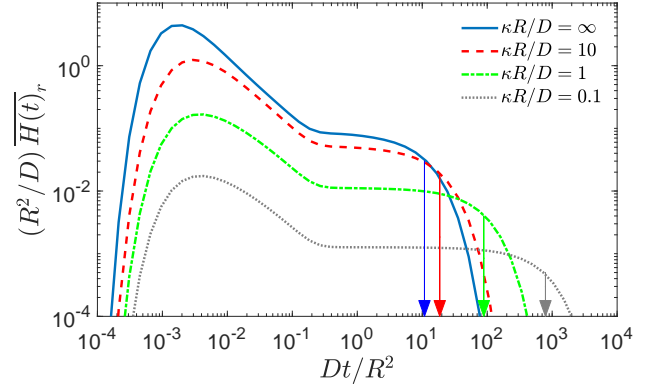}
\caption{Surface-averaged FPT density $\overline{H(t)}_r$ as function of $t$
for imperfect reactions, with $L/R=\pi$, $\rho/R=0.1$, $r/R=0.2$, and
$\varepsilon=0.2$.  All curves are obtained by numerical Laplace
inversion of \eqref{eq:H_tilde}. Arrows indicate the MFPT
$D\overline{T}_r/R^2$: $10.82$ ($\kappa =\infty$), $18.60$ ($\kappa =
10$), $88.58$ ($\kappa R/D = 1$), and $788.36$ ($\kappa R/D = 0.1$).
Note that (at fixed $D$) the MFPT grows with decreasing $\kappa$. In
this regime the MFPT becomes dominated by chemical reactivity,
$\overline{T}_r \sim 1/\kappa$ (see
Refs. \cite{Grebenkov17a,Grebenkov17}). The most probable FPT exhibits
a weak dependence on $\kappa$.  Note also the appearance of a
pronounced plateau-like region, which stretches over progressively
longer times scales upon lowering the reactivity $\kappa$. Hence,
there is a broad range of times with equiprobable realisations of the
FPT.  Length and time scales are fixed by setting $R=1$ and
$R^2/D=1$.}
\label{kappa}
\end{figure}

We also note that the probability density is broader in the case
$\ve=0.2$, with a flat intermediate region between the maximum hump
and the exponential cut-off (in the region $0.2\leq Dt/R^2\leq10$).
As the target size $\ve$ or the reactivity $\kappa$ decrease, the MFPT
increases and thus the exponential cut-off moves towards longer
times. In turn, the position and shape of the maximum remain
approximately constant (dominated by the initial distance and the
diffusivity $D$) so that the intermediate region expands, as we
checked for $\ve=0.05$ and for $\kappa R/D=0.1$ (not shown).  This is
a striking result: if the particle does not manage to find the target
and react within short times comparable to $t_m$ (around the maximum),
it explores the entire confining domain with eventual returns to the
target.  As a consequence, its reaction time is distributed almost
uniformly over a very broad range of times, up to the exponential
cut-off which is essentially determined by the MFPT.  The latter, in
turn, is dominated by the chemical reactivity, while the diffusive
search for the target provides only a sub-dominant contribution
\cite{Grebenkov17a} (see also \cite{Grebenkov17} for a general
discussion).  One can see that the low reactivity $\kappa$ leads to an
homogenisation of the search process, as evidenced in
Fig. \ref{kappa}, and the plateau-like region past the most probable
FPT extends over progressively longer scales when $\kappa$ becomes
smaller. As a consequence, the values of FPTs ranging over several
orders of magnitude appear to be almost equally probable.

Figure~\ref{fig:FPT3_kappa1}(b) shows that the volume-averaged FPT density
$\overline{H(t)}$ remains almost constant at short times and then has an
exponential cut-off. This almost uniform behaviour at short times resembles that
shown in Fig.~\ref{fig:FPT3_kappa1}(a). The only difference is that there is no
maximum at short times as some particles start infinitely close to the target.

\subsection{Biological implications}
\label{sec:application}

The function of biological cells to large extent relies on the passive diffusion
of regulatory molecules. In particular, the expression level of any gene is
controlled by the binding of transcription factor proteins. Inside the chromosome
a transcription factor locates its specific binding site via facilitated
diffusion combining volume search with one-dimensional sliding along the DNA,
as well as intersegmental jumps \cite{bvh,Elf07,gijs}. As many bacteria cells such
as the well studied {\it E.coli} or {\it bacilli} have distinct cylindrical shapes, the
analysis here provides an answer to the question how fast a given transcription
factor can reach the chromosome from the cytoplasm of the cell in the first place.
Our results demonstrate that for all considered scenarios the FPT to the nucleoid
is broadly distributed and may deviate significantly from the respective MFPT. For
reliable regulation it may thus be advantageous that transcription
factors, which often occur in very low copy numbers in a cell, are inhomogeneously
distributed in the cell \cite{kepes,kuhlman}, and may thus be kept close to their
target site on the DNA. This reasoning is in accord with results for the downstream
gene regulation model in \cite{otto} supporting the rapid search hypothesis
\cite{kolesov} as well as the geometry controlled few-encounter scenario of
\cite{Godec16}.

We also mention another relevant system for the cylindrical geometry,
namely, axons, the up to a meter long protrusions of neuronal cells,
whose diameter may span from 0.1 $\mu$m up to 20 $\mu$m
\cite{squire}. In the giant squid the diameter may even reach the
macroscopic size of 1 mm. In such an axon motor proteins detach from
the central bundle of microtubules, along which the motors actively
transport cargo. The motors' reattachment dynamics after unbinding,
governed by the results derived herein for imperfect reactions, have
shown to be important for the observed L{\'e}vy walk transport
\cite{granick}.

\section{Conclusion}

Although the necessity of knowing the full FPT distribution,
especially in situations when several length scales are involved, has
been emphasised earlier (e.g., in
\cite{Benichou11,Benichou14,Godec16,Godec16a}), not much progress has
been achieved in this direction.  For the first time, we discuss here,
using an analytical solution, the forms of the full first-passage time
distribution for different initial conditions in a cylindrical-annulus
geometry relevant for bacteria cells and neuronal axons. Due to the
quite large number of parameters in the system, the full distribution
of the FPT has a complicated structure and appreciably changes its
shape when the parameters are changed.  It would therefore be naive to
expect that the full complexity of the behaviour in the system could
be exhaustively characterised by just the first moment of this
distribution -- the mean first-passage time -- on which the previous
research has concentrated almost exclusively.

Within a self-consistent approach, proposed originally in a completely
different context in \cite{Shoup81}, we found explicit, approximate
expressions for the full FPT distribution, which we validated by
extensive numerical analysis.  One of the main features that we
uncovered is that, indeed, the full distribution has an important
structure and is rather sensitive to a slight variation of the
system's parameters.  Next, we showed that the MFPT turns out to be
several orders of magnitude longer than the most likely FPT, the
decisive quantity indicating when typically the first molecule arrives
at the target and triggers biochemical followup reactions. Therefore,
while the knowledge of the MFPT is certainly helpful and important, it
carries the danger of being misleading, given that the MFPT largely
overestimates the typical time scales involved in cellular
processes. In this context, former theoretical works devoted to the
minimisation of the MFPT do not necessarily reveal the optimal
conditions for the function of biological systems, because they do not
affect the most likely FPT.

Equally significant result of our analysis is the occurrence of an
extended plateau of the FPT distribution for lower reactivity
constants $\kappa$, signifying that over more than a decade all FPTs
within this range become equally probable and thus the triggering
events even more unfocused.  Moreover, within the unique geometric
setting, we could unveil intriguing dimensionality features of the
diffusive search in unbounded domains, for which the MFPT is infinite
and thus useless.  The derived asymptotic formulas correctly describe
intermediate regimes of the FPT distribution in the bounded case as
well.

Having available expressions for the full FPT distribution
will allow a more faithful evaluation of measured reaction dynamics
but also the planning of new experiments, in particular, when single
molecule resolution is accessible. We expect that our results will
lead to a new level of quantitative understanding of molecular
regulation processes on microscopic levels, for instance, a
renormalisation of rate constants extracted from MFPT interpretations.

\begin{acknowledgments}
DSG acknowledges the support under Grant No. ANR-13-JSV5-0006-01 of
the French National Research Agency.  RM aknowledges support from the
Deutsche Forschungsgemeinschaft through projects ME 1535/6-1 and ME
1535/7-1, as well as from the Foundation for Polish Science within an
Alexander von Humboldt Polish Honorary Research Fellowship is
gratefully acknowledged.
\end{acknowledgments}

\newpage ~ \newpage
%\begin{appendix}

%%%%%%%%%% Merge with supplemental materials %%%%%%%%%%
%%%%%%%%%% Prefix a "S" to all equations, figures, tables and reset the counter %%%%%%%%%%
\setcounter{equation}{0}
\setcounter{section}{0}
\setcounter{figure}{0}
\setcounter{table}{0}
\setcounter{page}{1}
\makeatletter
\renewcommand{\theequation}{S\arabic{equation}}
\renewcommand{\thefigure}{S\arabic{figure}}
\renewcommand{\thesection}{\Roman{section}}

%\onecolumn
\onecolumngrid

\noindent
\textbf{\textsf{\Large Supplemental Information for the article ``\@title''}}

%\textbf{\textsf{\Large Supplemental Information for the article ``Towards a 
%full quantitative description of single-molecule reaction
%kinetics in biological cells''}}

\section{Determination of the expansion coefficients}
\label{coeffs}

To compute the coefficients $a_n$ within the self-consistent
approximation (SCA), we first substitute $\tilde S(r,\theta; p)$
defined in \eqref{eq:ansatz} into the modified boundary condition in
\eqref{eq:BC_modified}, which gives
\begin{equation} \label{eq:auxil21}
u'_0(\rho) + \sum\limits_{n=0}^\infty a_n \, g'_n(\rho) \cos(n\theta) = \frac{Q}{R^2}\, \Theta(\ve - \theta) ,
\end{equation}
where
\begin{equation}  \label{eq:gnp_free}
g'_n(r) = \frac{\alpha_n}{L} \bigl[ I_1(\alpha_n r/L) K_1(\alpha_n R/L) - K_1(\alpha_n r/L) I_1(\alpha_n R/L) \bigr]. 
\end{equation}
Multiplying \eqref{eq:auxil21} by $\cos(n\theta)$ (with $n =
0,1,2,\ldots$) and integrating over $\theta$ from $0$ to $\pi$, one
gets
\begin{eqnarray}
Q &=& \frac{\pi R^2}{\ve} \bigl( u'_0(\rho) + a_0 g'_0(\rho) \bigr),  \\ 
a_n &=& \frac{2Q}{\pi R^2 g'_n(\rho)} \, \frac{\sin n\ve}{n} \quad (n = 1,2,\ldots), 
\end{eqnarray}
so that
\begin{equation}
a_n = 2 \, \frac{u'_0(\rho) + a_0 g'_0(\rho)}{g'_n(\rho)}\, \frac{\sin n\ve}{n\ve} \quad (n = 1,2,\ldots).
\end{equation}
Consequently, $ \tilde S(r,\theta; p)$ writes
\begin{equation} \label{eq:t_solution}
\begin{split}
& \tilde S(r,\theta; p) = \frac{R^2}{D} \biggl( u_0(r) + a_0 g_0(r) 
 + 2 \bigl( u'_0(\rho) + a_0 g'_0(\rho)\bigr)  
\sum\limits_{n=1}^\infty  \frac{g_n(r)}{g'_n(\rho)}  \, \frac{\sin n\ve}{n\ve } \, \cos(n\theta) \biggr). \\
\end{split}
\end{equation}
The coefficient $a_0$ (and hence, the trial current $Q$) remains a
free parameter which is to be chosen in a self-consistent way.  Within
the SCA proposed in \cite{Shoup81} (see also
\cite{Grebenkov17,Grebenkov17a} for more details on the adaption of
this scheme to the first-passage phenomena) the closure relation is
obtained by requiring that the mixed boundary condition in
\eqref{eq:BC_original} holds not locally but \textit{on average}.

For this purpose, we substitute this expression into
\eqref{eq:BC_original} and integrate the resulting expression over
$\theta$ from $0$ to $\ve$ to get
\begin{equation}  \label{eq:a0}
a_0 = \frac{1-\eta}{s \, g_0(\rho)} \,, 
\end{equation}
with $\eta$ defined by \eqref{eq:eta}.  Combining these results, we
get an approximate but explicit solution
\begin{eqnarray}
\label{eq:t_solution2}
\tilde{S}(r,\theta;p)= \frac{1}{p}\biggl(1 - \eta \, \frac{g_0(r)}{g_0(\rho)}+2\eta\, \frac{g'_0(\rho)}{g_0(
\rho)}\sum\limits_{n=1}^\infty\frac{g_n(r)}{g'_n(\rho)}\frac{\sin n\ve}{n\ve}
\cos(n\theta) \biggr) ,
\end{eqnarray}
from which \eqref{eq:H_tilde} follows.

\section{Numerical validation of the results obtained within the SCA}
\label{sec:validation}

The proposed SCA provides an exact solution of the modified boundary
value problem, in which the mixed Robin-Neumann boundary
condition in \eqref{eq:BC_original} is substituted by the effective
inhomogeneous Neumann condition in \eqref{eq:BC_modified}.  To check
the accuracy of this approximation and hence, of the resulting FPT
distribution, we solve the original modified Helmholtz equation by a finite
elements method (FEM) implemented in Matlab's PDEtool. Setting
$\tilde{H}(r,\theta;p) = u(r/R,\theta;pR^2/D)$ we rewrite the
modified Helmholtz equation in cylindrical coordinates as
\begin{equation}  \label{eq:Helm1}
(\partial/\partial r)r(\partial/\partial r)u+r\pi^2(R^2/L^2)(\partial^2/\partial\theta^2)u-rsu= 0, 
\end{equation}
which has to be solved on the rectangular domain
$(\rho/R,1)\times(0,\pi)$ with Neumann boundary condition imposed
everywhere except for the segment $(\rho)\times(0,\ve)$, for which
\eqref{eq:BC_original} reads
\begin{equation}  \label{eq:Helm2}
\biggl(u - \frac{D}{R\kappa} \, \frac{\partial u}{\partial r}\biggr)_{|r=\rho} = 1 \qquad (0 < \theta < \ve).
\end{equation}
Once the solution $u$ is found on mesh vertices, one can also compute
its volume and surface averages.  The volume-averaged quantity
$\overline{\tilde{H}(p)}$ is obtained by a numerical integration of
the solution over the computation domain, whereas the surface-averaged
quantity $\overline{\tilde{H}(p)}_r$ is evaluated by first a linear
interpolation of the solution to the vertical line at $r$ and then by a
numerical integration over this line.

The accuracy of the numerical solution of this problem and its
averages is controlled by the maximal mesh size $h$, i.e., the
largest allowed diameter of triangles of the mesh used to discretise
the computational domain.  In particular, the maximal mesh size should
be much smaller than the length $\ve$ of the reactive segment.  This
condition limits the accessible target heights $\ve$.  In our
numerical analysis, we set $h = 0.005$ that results in meshes with
more than 200~000 triangles.  It should be noted that since the
numerical solution should be repeated for many values of $p$ (or $s$),
computations with even larger meshes (and thus smaller $h$ and
$\ve$) become too time-consuming.  Moreover, this long computation
also prohibits using the Talbot algorithm for the Laplace inversion.
Alternatively, the probability density $H(r,\theta;t)$ in time domain
might be computed by solving directly the diffusion equation for the
survival probability, but this solution would be even more
time-consuming and limited to a relatively narrow range of times.  For
these reasons, we focus here on a numerical validation of the SCA only
in the Laplace domain, i.e., checking the moment-generating function
over a wide range of variation of the parameter $p$ of the Laplace
transform, instead of the probability density function itself.  From a
formal viewpoint, this is equivalent to validating the FPT
distribution since both quantities are uniquely linked by the Laplace
transform.

\begin{figure*}
\includegraphics[width=85mm]{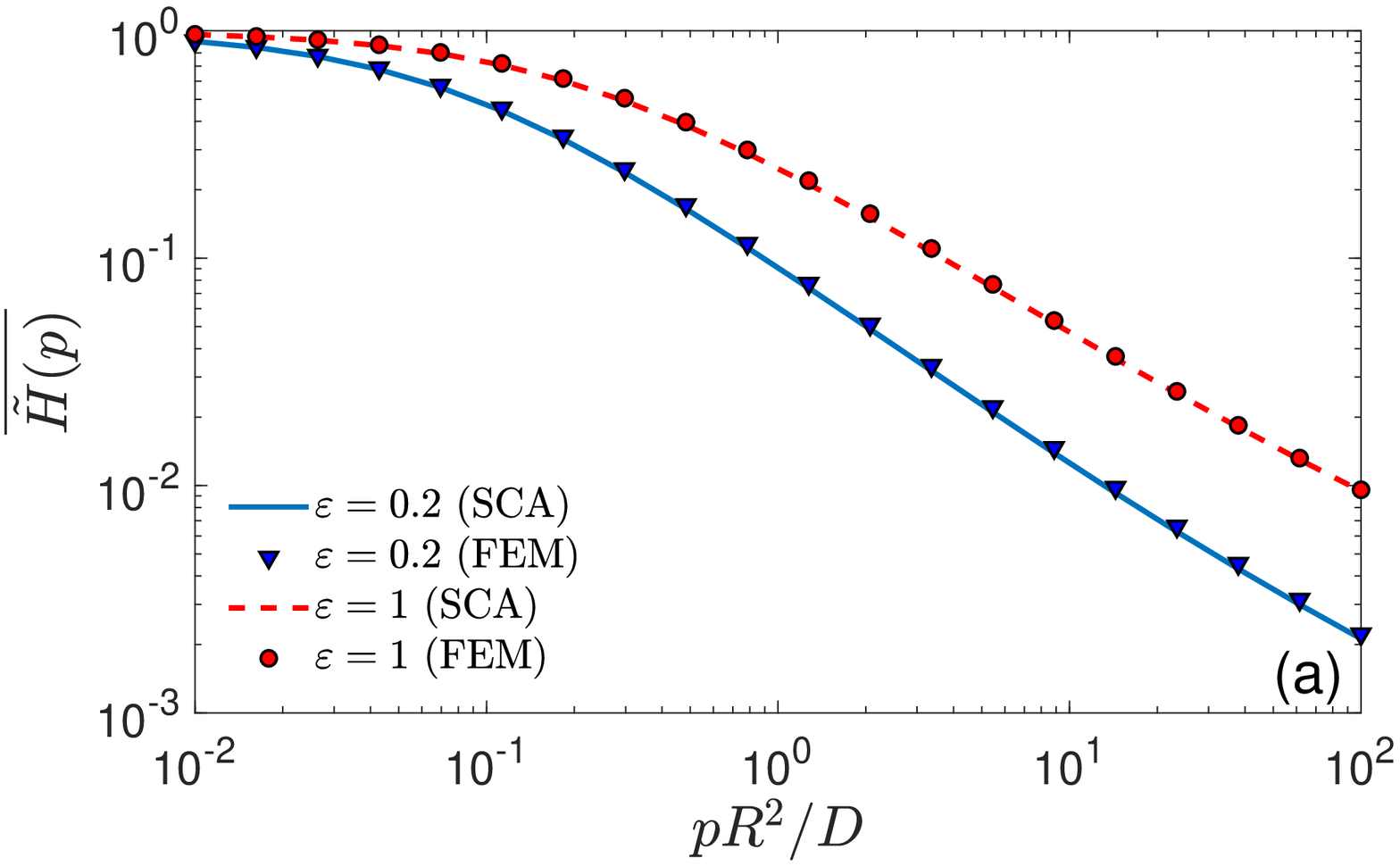}
\includegraphics[width=85mm]{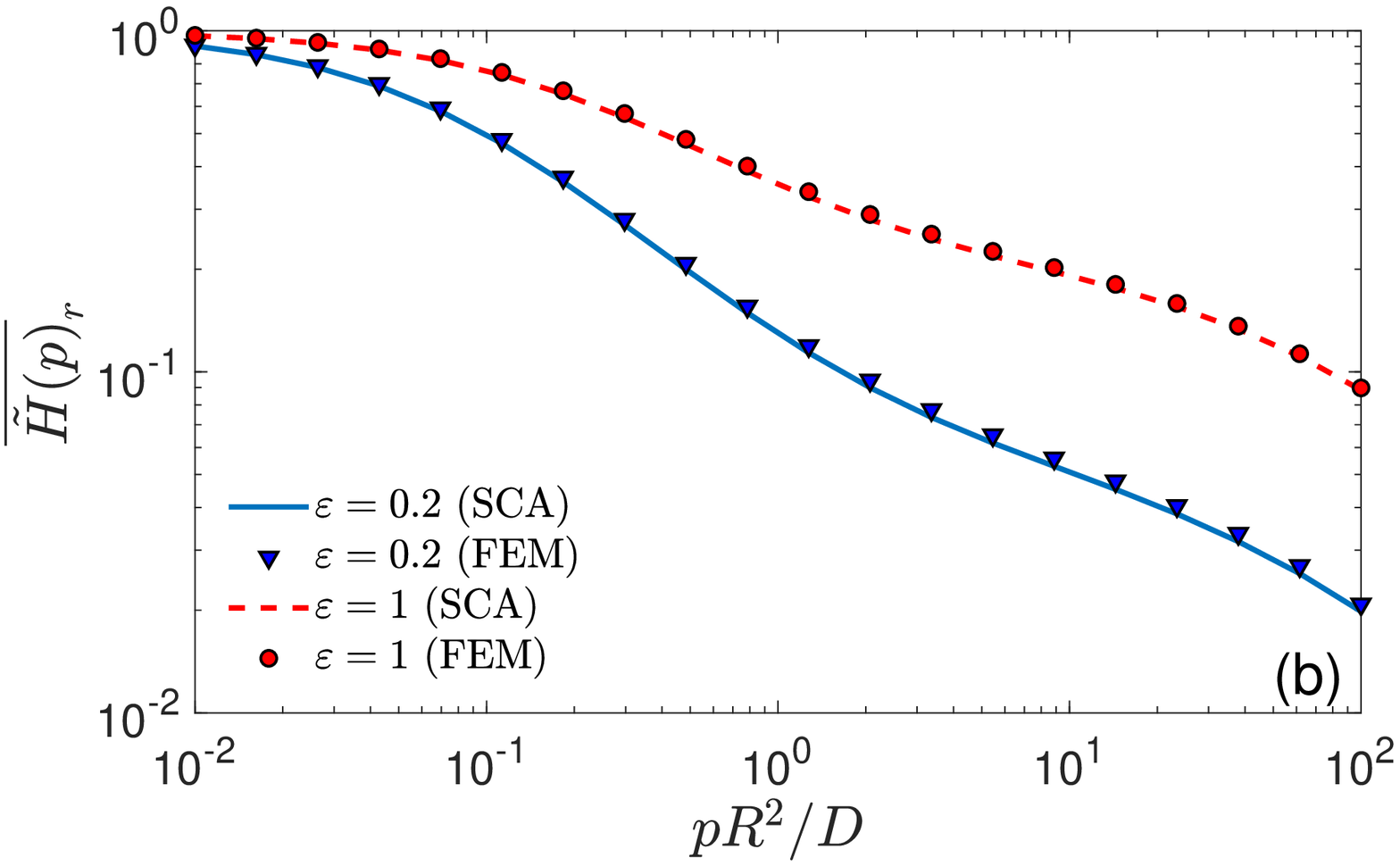}
\caption{
Volume-averaged {\bf (a)} and surface-averaged {\bf (b)}
Laplace-transformed FPT densities $\overline{\tilde{H}(p)}$ and
$\overline{\tilde{H}(p)}_r$ as function of $s = pR^2/D$ for $L/R =
\pi$, $\rho/R = 0.1$, $r/R = 0.2$, $\kappa=\infty$, and two values of
$\varepsilon$ indicated in the plot.  Lines show \eqref{lap_ex} {\bf
(a)} and \eqref{eq:H_tilde_surf} {\bf (b)} of our self-consistent
approximation, in which the series in \eqref{eq:Rve} is truncated to
$N = 100$ terms.  Symbols present a FEM numerical solution of the
modified Helmholtz \eqref{eq:Helm1}, with mixed boundary condition
\eqref{eq:Helm2} with the maximal mesh size $h = 0.005$.
Length and time scales are fixed by setting $R = 1$ and $R^2/D=1$.}
\label{fig:FEM_Hr_eps}
% A_Oshanin3_FEM_fignew2(0);
% A_Oshanin3_FEM_fignew2(1);
\end{figure*}

Figure \ref{fig:FEM_Hr_eps} shows excellent agreement between the
result based on the SCA and the FEM solution of the original mixed
boundary value problem for $\ve=0.2$ and $\ve=1$ (we do not consider
the case $\ve = \pi$ for which the SCA yields the exact solution, see
Section \ref{sec:eps_pi} of the SI).  Moreover, the result of the SCA
converges rapidly as the upper summation truncation $N$ of the series
in \eqref{eq:Rve} increases.  In particular, the results for $N = 50$
(not shown) and $N = 100$ are barely distinguishable.  Small
deviations at large $p$ can be attributed to (i) inaccuracy of the FEM
solution (and the consequent numerical integrations for getting volume
and surface averages), and (ii) intrinsic small differences between
the original and modified problems.  Nevertheless, the quality of the
SCA is quite impressive.

\begin{figure*}
\includegraphics[width=85mm]{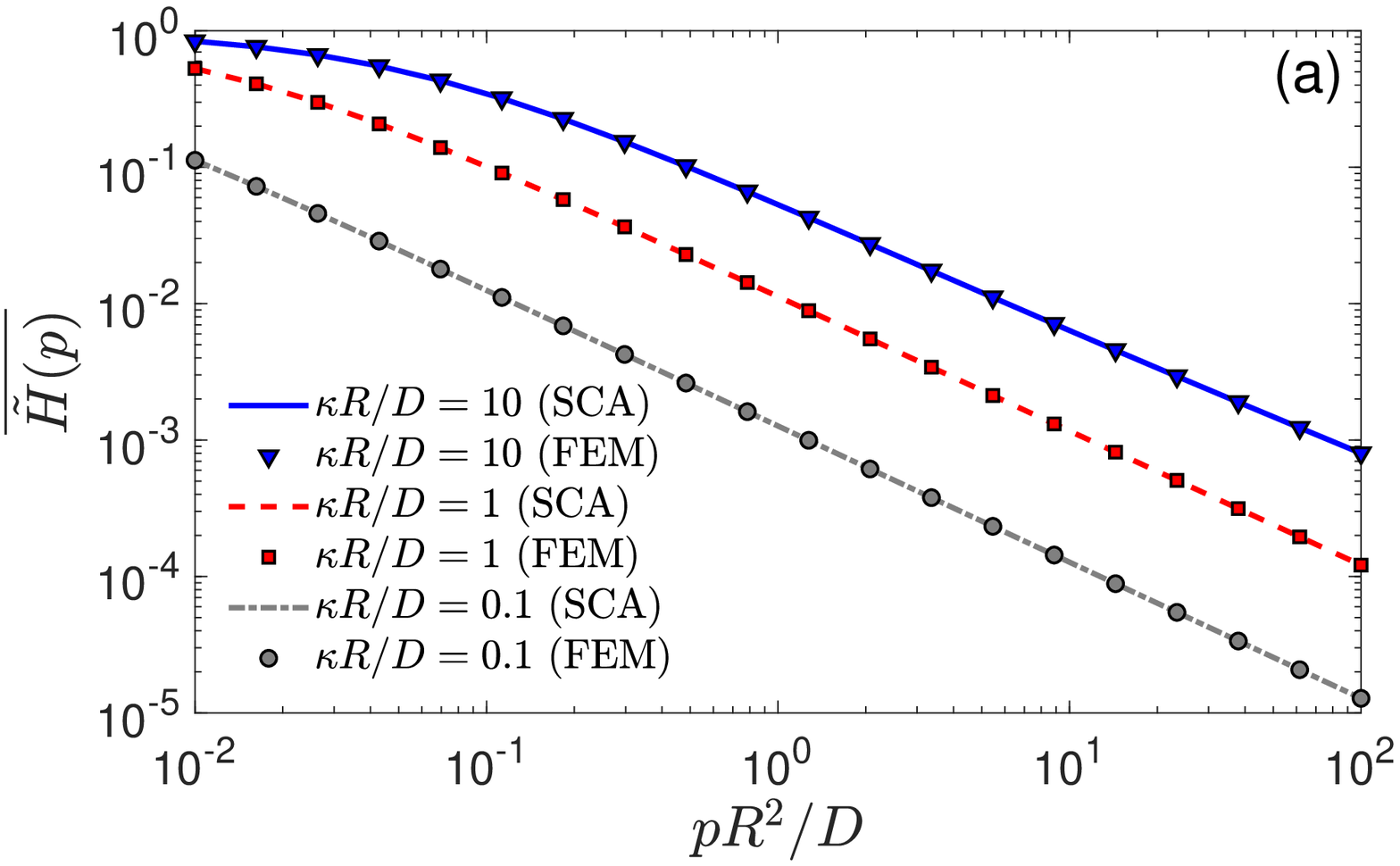}
\includegraphics[width=85mm]{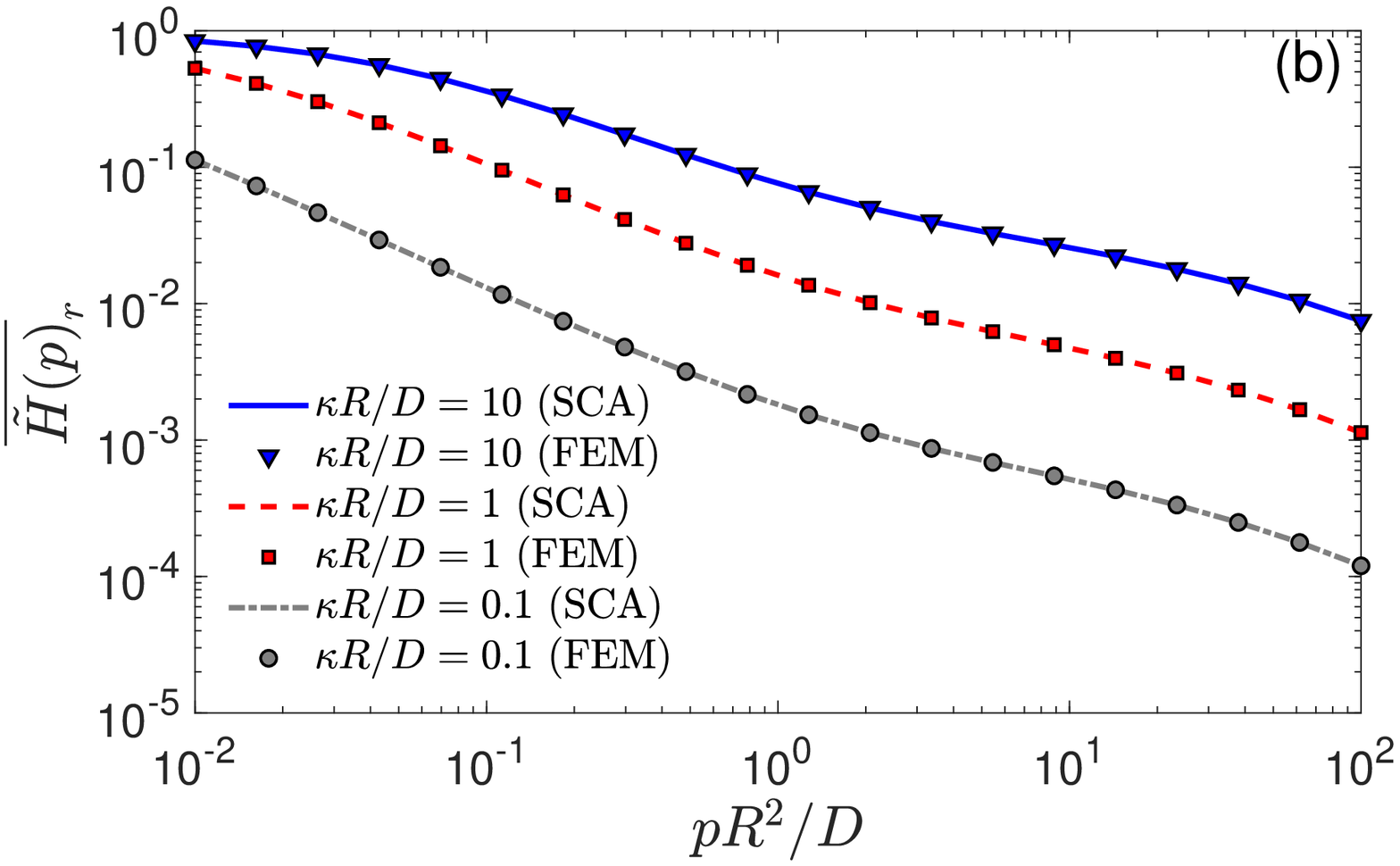}
\caption{Volume-averaged {\bf (a)} and surface-averaged {\bf (b)}
Laplace-transformed FPT densities $\overline{\tilde{H}(p)}$ and
$\overline{\tilde{H}(p)}_r$ as function of $s = pR^2/D$ for $L/R =
\pi$, $\rho/R = 0.1$, $r/R = 0.2$, $\varepsilon = 0.2$, and three
values of $\kappa$, indicated in the plot.  Lines show \eqref{lap_ex}
{\bf (a)} and \eqref{eq:H_tilde_surf} {\bf (b)} of our self-consistent
approximation, in which the series in \eqref{eq:Rve} is truncated to
$N = 100$ terms.  Symbols present a FEM numerical solution of the
modified Helmholtz \eqref{eq:Helm1}, with mixed boundary condition
\eqref{eq:Helm2} with the maximal mesh size $h = 0.005$.
Length and time scales are fixed by setting $R = 1$ and $R^2/D=1$.}
\label{fig:FEM_Hr_kappa}
% A_Oshanin3_FEM_fignew3(0);
% A_Oshanin3_FEM_fignew3(1);
\end{figure*}

The SCA becomes particularly robust for imperfect reactions on the
target.  Figure \ref{fig:FEM_Hr_kappa} compares the SCA to the FEM
solution for $\ve = 0.2$ and several values of $\kappa$.  The SCA
predictions and FEM solutions are indistinguishable at the logarithmic
scale.  Lastly, as discussed in Ref. \cite{Grebenkov17}, the SCA is
generally getting more accurate for smaller target size $\ve$, smaller
reactivity $\kappa$, and for starting points that are not too close to
the target. Moreover, it was shown recently in Ref.~\cite{diet}, in
which the self-propulsion velocity of catalytically-active colloids
was studied by a similar method, that also for an arbitrary $\ve$ the
SCA provides an accurate description and only slightly overestimates
the numerical factors.

\section{The probability density in time domain}
\label{sec:time}

The Laplace-transformed probability density $\tilde{H}(r,\theta;p)$
and the related surface and volume-averaged quantities can be inverted
by using the residue theorem.  For this purpose, one searches for the
poles of $\tilde{H}(r,\theta;p)$ in the complex plane $p\in {\mathbb
C}$.  In general, there are two groups of poles, which can be
identified as (i) zeros of functions $g'_n(\rho)$ and (ii) zeros of
the function $g_0(\rho)/\eta$.

(i) Introducing the auxiliary function
\begin{eqnarray}   \label{eq:hatG_aux}
\hat{G}(z) \equiv -iz\biggl(I_1(-iz\rho/R) K_1(-iz) - K_1(-iz\rho/R) I_1(-iz)\biggr) 
= - \frac{\pi i}{2} z \biggl(Y_1(z\rho/R) J_1(z) - J_1(z\rho/R) Y_1(z)\biggr) ,
\end{eqnarray}
one writes $g'_n(\rho) = \hat{G}(i\alpha_n R/L)/R$, where $\alpha_n =
\sqrt{\pi^2 n^2 + pL^2/D}$.  One can check that the function
$\hat{G}(z)$ has infinitely many zeros all lying on the real axis.
Since the Bessel functions obey
\begin{equation*}
\begin{array}{l} 
J_n(-x) = (-1)^n J_n(x) \\  
Y_n(-x) = (-1)^n \bigl(Y_n(x) + 2i J_n(x)\bigr) \\  \end{array}  \qquad (x > 0),
\end{equation*}
one deduces that $\hat{G}(-z) = - \hat{G}(z)$.  We can thus focus only
on the positive zeros of $\hat{G}(z)$ denoted as $\hat{z}_k$ (with $k
= 1,2,\ldots$).  We relate them to the zeros of $g'_n(\rho)$:
\begin{equation} 
p_{n,k} = - D \biggl(\frac{\hat{z}_k^2}{R^2} + \frac{\pi^2 n^2}{L^2}\biggr), \qquad
\left(\begin{array}{l} n=1,2,\ldots \\ k = 1,2,\ldots \end{array}\right). 
\end{equation} 
For each $n$, all the poles $p_{n,1}, p_{n,2}, \ldots$ are expected to
be simple.  In turn, it is possible to tune $R$ and $L$ to make some
poles with different $n$ coincide and thus be of the order higher than
$1$.  At such poles, the computation of the residue would be more
involved.

(ii) Similarly, we introduce the function
\begin{eqnarray}  
G_r(z) \equiv I_0(-iz r/R) K_1(-iz) + K_0(-iz r/R) I_1(-iz) 
= \frac{\pi i}{2} \biggl(Y_0(z r/R) J_1(z) - J_0(z r/R) Y_1(z)\biggr) ,
\end{eqnarray}
to write $g_n(r) = G_r(i\alpha_n R/L)$.  Note that this function is
also antisymmetric: $G_r(-z) = -G_r(z)$.  The poles of the function
\begin{equation*}
\frac{\eta}{g_0(\rho)} = \frac{1}{g_0(\rho) - \bigl(\frac{\pi D}{\kappa\ve} + \frac{L}{\pi} \Rve_\ve\bigr) g'_0(\rho)}  
\end{equation*}
are related to the zeros of the function
\begin{equation}
F(z) \equiv G_\rho(z) - \biggl(\frac{\pi D}{R \kappa\ve} + \frac{L}{\pi R} \Rve_\ve(z)\biggr) \hat{G}(z) ,
\end{equation}
where
\begin{equation}
\Rve_\ve(z) = - \frac{2\pi R}{L} \sum\limits_{n=1}^\infty \biggl(\frac{\sin n\ve}{n\ve}\biggr)^2 \,
\frac{G_\rho\bigl(\sqrt{z^2 - \pi^2 n^2 R^2/L^2}\bigr)}{\hat{G}\bigl(\sqrt{z^2 - \pi^2 n^2 R^2/L^2}\bigr)} \,.
\end{equation}
Denoting the positive zeros of $F(z)$ as $z_k$, we get the poles as
\begin{equation}
p_{0,k} = - \frac{D}{R^2} z_k^2  \qquad (k=1,2,\ldots).
\end{equation}
Having identified all the poles, one can formally invert the Laplace
transform by applying the residue theorem.

\section{Long and short time asymptotical behaviour of the FPT density}
\label{long_short}

We here summarise the relations relevant for the calculation of the
asymptotic behaviours of the FPT density.

\subsection*{Long-time behaviour}

With $p\to0$ we also have $s\to0$, which then leads to $\alpha_n=\pi n+O(s)$
for $n>0$ and $\alpha_0=\sqrt{s} \, L/R$. We get
\begin{equation}
g_0(r)\simeq\frac{1}{\sqrt{s}}\biggl(1+s\biggl(\frac{r^2}{4R^2}-\frac12\ln(r/R)
\biggr)+O(s^2)\biggr),
\end{equation}
so that 
\begin{eqnarray}
\frac{g_0(r)}{g_0(\rho)}&\simeq&1-\biggl(\frac{\rho^2-r^2}{4R^2}+\frac{\ln(r/
\rho)}{2}\biggr)s+O(s^2),\\
\frac{g'_0(\rho)}{g_0(\rho)}&\simeq&-s\frac{1-(\rho/R)^2}{2\rho}+O(s^2),\\
\Rve_\ve&\simeq&\Rve_\ve(p=0)+O(s),\\
\label{eq:eta_p0}
\eta^{-1}&\simeq&1+p\overline{T}_\rho+O(p^2),
\end{eqnarray}
where $\overline{T}_\rho$ from \eqref{surftau} is the
surface-averaged MFPT studied in Ref. \cite{Grebenkov17a}.

\subsection*{Short-time behaviour}

The short-time behaviour corresponds to the limit $p\to\infty$, in
which we get
\begin{eqnarray}
\label{eq:auxil10}
\frac{g_n(\rho)}{g'_n(\rho)} \simeq -\frac{L}{\alpha_n}\frac{K_0(\alpha_n
\rho/L)}{K_1(\alpha_n \rho/L)}
\simeq -\frac{L}{\alpha_n}\biggl(1-\frac{1}{2\alpha_n\rho/L}+\frac{3}{8(
\alpha_n\rho/L)^2}+\ldots\biggr) 
\end{eqnarray}
and
\begin{equation}
\Rve_\ve\simeq2\pi\sum\limits_{n=1}^\infty\biggl(\frac{\sin n\ve}{n\ve}\biggr)^2
\biggl(\frac{1}{\alpha_n}-\frac{L}{2\rho\alpha_n^2}+\ldots\biggr).
\end{equation}
Setting $z=\sqrt{s}L/R$ we identify the leading part in the first term by writing
\begin{equation}  
\label{eq:auxil11}
\Rve_\ve\simeq2\pi\sum\limits_{n=1}^\infty\biggl(\frac{\sin n\ve}{n\ve}\biggr)^2
\biggl(\frac{1}{z}-\frac{\pi^2n^2}{z\alpha_n(\alpha_n+z)}-\frac{L}{2\rho\alpha_n
^2}+\ldots\biggr). 
\end{equation}
The first term yields the leading contribution equal to
$\frac{\pi(\pi-\ve)}{\ve}z^{-1}$, in virtue of the identity
\begin{equation}
\sum\limits_{n=1}^\infty\biggl(\frac{\sin n\ve}{n\ve}\biggr)^2=\frac{\pi-\ve}{2
\ve}.
\end{equation}
The second sum in \eqref{eq:auxil11} can be written as
\begin{equation}
\begin{split}
\Rve_\ve^{(2)}&=-\frac{\pi^3}{z\ve^2}\sum\limits_{n=1}^\infty\frac{1-\cos2n\ve}{
\pi^2n^2+z^2+\alpha_nz}
\simeq-\frac{\pi^3}{z\ve^2}\sum\limits_{n=1}^\infty\frac{1-\cos2n\ve}{\pi^2n^2
+2z^2},\\
\end{split}
\end{equation}
where we substituted $\alpha_n\simeq z$ (for large $z$) in the
denominator of the last term. Using the identity
\begin{equation}
\sum\limits_{n=1}^\infty\frac{\cos(\pi nx)}{\pi^2n^2+z^2}=\frac{\cosh(1-x)z}{2z
\sinh z}-\frac{1}{2z^2},
\end{equation}
which is valid for $0\leq x\leq1$, we find
\begin{equation}
\Rve_\ve^{(2)}\simeq-\frac{\pi^3}{z\ve^2}\frac{\cosh(z\sqrt{2})-\cosh(z\sqrt{2}(1
-2\ve/\pi))}{2\sqrt{2}z\sinh(z\sqrt{2})}.
\end{equation}
We emphasise that this relation is only valid for $\ve\leq\pi/2$. At large $s$
(or $z$), one thus gets a very accurate approximation
\begin{equation}
\label{eq:auxil12}
\Rve_\ve^{(2)}\simeq-\frac{\pi^3}{2\sqrt{2}z^2\ve^2}\bigl(1-e^{-2\sqrt{2}z\ve/
\pi}\bigr).
\end{equation}
Finally, in the third term of \eqref{eq:auxil11} we approximate again $\alpha_n^2
\approx z^2$, which yields $-\frac{\pi(\pi-\ve)L}{2\ve\rho}z^{-2}$. Combining these
results, we obtain
\begin{equation}
\Rve_\ve\simeq\frac{\pi(\pi-\ve)R}{\ve L}s^{-1/2}-\mathcal{A}s^{-1}+O(s^{
-\frac32}) ,
\end{equation}
with
\begin{equation}
\A =\frac{\pi R^2}{\ve L^2}\biggl(\frac{\pi^2}{2\sqrt{2}\ve}+\frac{(\pi-\ve)L}{2
\rho}\biggr),
\end{equation}
in which we neglected the exponential correction in \eqref{eq:auxil12}.

From \eqref{eq:auxil10}, we also get
\begin{equation}
- \frac{g'_0(\rho)}{g_0(\rho)}\simeq\frac{\sqrt{s}}{R}+\frac{1}{2\rho}-\frac{R}{
8\rho^2}s^{-1/2}+O(s^{-1}).
\end{equation}
Combining these results, we find the large $s$ asymptotic behaviour of
\eqref{eq:eta},
\begin{eqnarray}
\label{eq:eta_large}
\eta \simeq \frac{\ve}{\pi}\biggl[s^{1/2}\frac{D}{\kappa R}+\biggl(1+\frac{
D}{2\kappa\rho}\biggr)
-\biggl(\frac{DR}{8\kappa\rho^2} + \frac{\pi R}{\sqrt{8} \ve L}\biggr)
s^{-1/2}+O(s^{-1})\biggr]^{-1} .
\end{eqnarray}

When $\kappa=\infty$, \eqref{eq:eta_large} becomes
\begin{equation}
\label{eq:eta_pinf}
\eta\simeq\frac{\ve}{\pi}+\frac{\sqrt{D}}{2\sqrt{2}L}p^{-1/2}+O(p^{-1}).
\end{equation}

When the particles start from a cylindrical surface at $r$, to obtain
\eqref{eq:Ht_r_t0} we need the large-$s$ asymptotic relation
\begin{equation}
\label{large-s}
g_0(r)\simeq\frac{\cosh(\sqrt{s}(1-\frac{r}{R}))}{\sqrt{s}\sqrt{r/R}} \biggl(1 
-\frac{\tanh(\sqrt{s}(1-\frac{r}{R}))}{8}(3+R/r) s^{-1/2} +O(s^{-1}) \biggr),
\end{equation}
from which we find the Laplace-transformed FTP density
\begin{equation}
\label{eq:Hp_asympt}
\overline{\tilde H(p)}_r \simeq\eta\sqrt{\rho/r}\exp\bigl(-(r-\rho)\sqrt{p/D}\bigr)
\biggl(1 + \sqrt{D} \frac{\sqrt{R/\rho}-\sqrt{R/r}}{8R} \, p^{-1/2}+O(p^{-1})\biggr)
\end{equation}
for $r<R$ and $p\to\infty$.

For $\kappa<\infty$, \eqref{eq:eta_large} becomes
\begin{equation}
\label{eta_imp}
\eta \simeq\frac{\ve}{\pi}\frac{\kappa}{\sqrt{D}} \, p^{-1/2}-\frac{\ve}{\pi}
\biggl(\frac{\kappa^2}{D}+\frac{\kappa}{2\rho}\biggr)\, p^{-1}+O(p^{-\frac32}),
\end{equation}
which leads to \eqref{imperfect}.

\begin{figure}
\includegraphics[width=85mm]{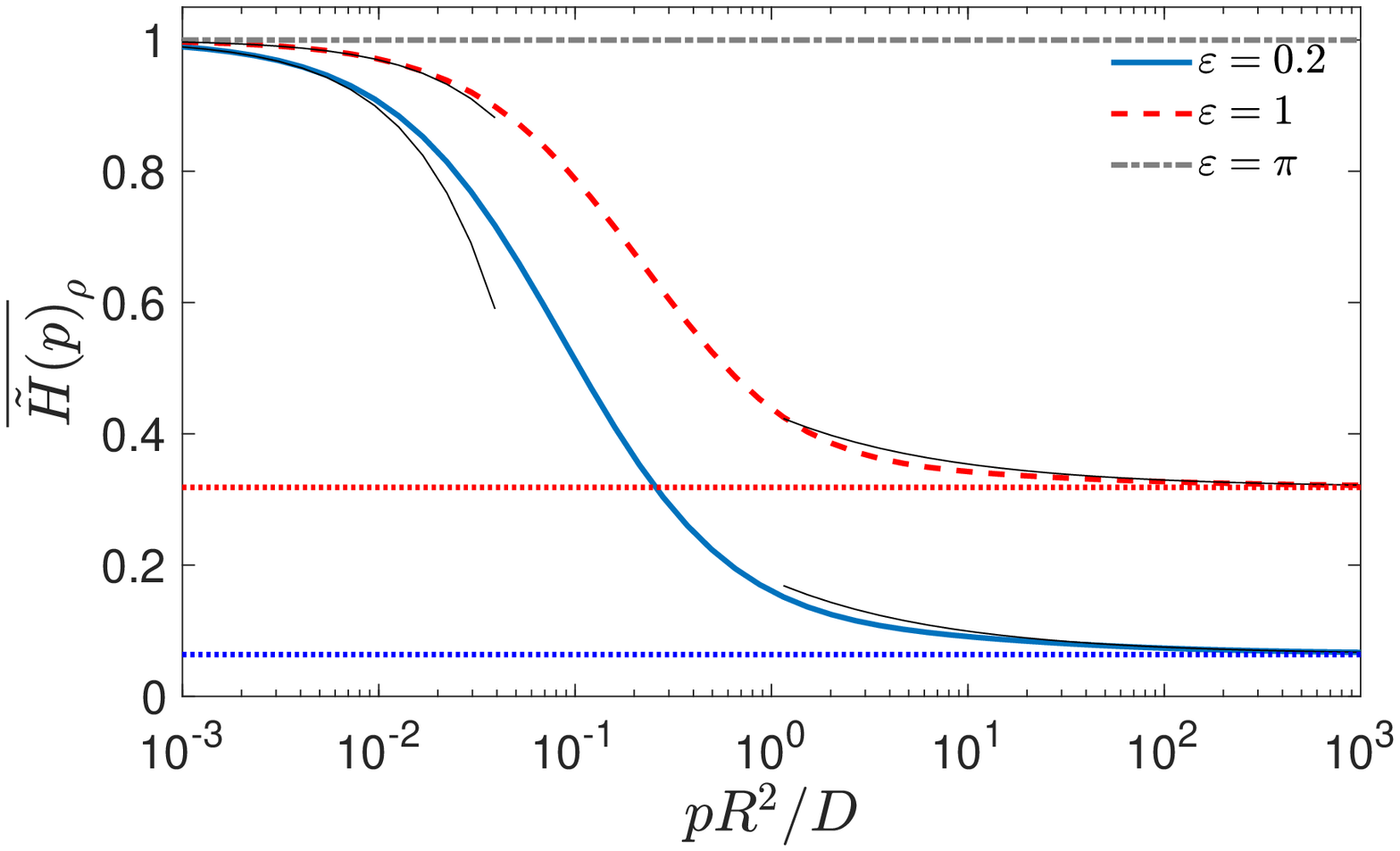}
\caption{Surface-averaged Laplace-transformed FPT density $\overline{\tilde{
H}(p)}_\rho$ as function of $s=pR^2/D$ for $L/R = \pi$, $\rho/R =
0.1$, $\kappa=\infty$, and the three values of $\varepsilon$ indicated
in the plot. Thin solid lines show the asymptotic
\eqref{eq:eta_p0} and \eqref{eq:eta_pinf} as $s\to0$ and $s\to\infty$,
respectively. Dotted horizontal lines indicate the asymptotic limit
$\ve/\pi$ of the FPT density as $s\to\infty$. Length and
time scales are fixed by setting $R = 1$ and $R^2/D = 1$.}
\label{fig:Hrho_p}
\end{figure}

Figure~\ref{fig:Hrho_p} illustrates the behaviour of the
surface-averaged probability density $\overline{\tilde{H}(p) }_\rho$
for three target heights $\epsilon$ in the case of perfect reactions
($\kappa=\infty$).  We observe that $\overline{\tilde{H}(p) }_\rho$
linearly approaches unity as $p\to 0$ (see \eqref{eq:eta_p0}) and
reaches a constant $\ve/\pi$, that is, the fraction of the target
area, as $p\to\infty$. This is a consequence of the uniform surface
average: particles that start from the target are immediately absorbed
and thus not affected by diffusion-reaction processes. The case $\ve=
\pi$ corresponds to the fully absorbing inner cylinder, with $\overline{\tilde{
H}(p)}_\rho=\eta=1$.

\section{Target on the whole inner cylinder}
\label{sec:eps_pi}

We consider the special case of the target extended to the whole inner
cylinder (i.e., $\ve = \pi$), for which $a_n = 0$ ($n > 0$), $\Rve_\pi
= 0$, and thus the SCA yields
\begin{equation}  \label{eq:Stilde_pi}
\tilde S(r,\theta;p) = \frac{1}{p} \biggl(1 - \eta \, \frac{g_0(r)}{g_0(\rho)}\biggr) ,
\end{equation}
with
\begin{equation}  \label{eq:eta_pi}
\eta = \biggl(1 - \frac{g'_0(\rho)}{g_0(\rho)} \, \frac{D}{\kappa}\biggr)^{-1} .
\end{equation}
One can easily check that this is the {\it exact} solution of the
original problem with a partially absorbing inner cylinder, i.e., this
function satisfies the Robin boundary condition at $r = \rho$:
\begin{equation}
\bigl(D \partial_n \tilde S + \kappa \tilde S\bigr)_{r = \rho} = 0 .
\end{equation}
In other words, our approximation becomes exact in the case $\ve =
\pi$.

From \eqref{eq:Stilde_pi}, we also get the MFPT to the inner cylinder:
\begin{equation}  \label{eq:MFPT_whole}
{T} = \frac{R^2-\rho^2}{2\rho \kappa} + \frac{\rho^2 - r^2}{4D} + \frac{R^2 \ln(r/\rho)}{2D}  \,,
\end{equation}  
whereas its volume average reads
\begin{equation} \label{eq:GMFPT_whole}
\overline{T} = \frac{R^2-\rho^2}{2\rho \kappa} +
\frac{R^2}{8D} \biggl(\frac{4\ln(R/\rho)}{1-(\rho/R)^2} - 3 + (\rho/R)^2\biggr).
\end{equation}

The above exact solution suggests an interesting interpretation of the
coefficient $\Rve_\ve$ for the general case of an arbitrary $\ve$.  In
fact, when the reactive region is only a part of the inner cylinder
(i.e., $\ve < \pi$), a diffusing particle will undergo multiple
reflections by the remaining part of the inner cylinder until it
reaches the target.  If the target was composed of many small regions
uniformly distributed over the inner cylinder, such a partially
reactive boundary could be described by the Robin boundary condition
over the whole inner cylinder, with an effective reactivity
$\kappa_{\rm eff}$.  In our case, the target is a single absorbing
region, so that this homogenisation argument is {\it a priori} not
applicable.  Nevertheless, one can still introduce an effective,
\textit{apparent} reactivity by looking at the form of $\eta$ in
\eqref{eq:eta}:
\begin{equation}  \label{eq:kappa_eff}
\frac{\pi}{\kappa \ve} + \frac{L}{\pi D} \Rve_\ve = \frac{1}{\kappa_{\rm eff}} \,.
\end{equation}
This relation can be thought of as an extension of the celebrated
Collins-Kimball relations for the system under study (see
\cite{Grebenkov17,Grebenkov17a} for more details).

When the target is perfectly reactive, $\kappa = \infty$ and thus the
effective reactivity represents the effect of the mixed
Dirichlet-Neumann boundary condition: $\kappa_{\rm eff} =
\frac{D\pi}{L\Rve_\ve}$.  If in addition the target is partially
reactive, the overall reactivity is further reduced, as the particle
has to reach the reactive part and then overcome the energy barrier.
Interpreting the inverse of the reactivity as a ``resistance'',
\eqref{eq:kappa_eff} implies that the respective resistances
enter additively, precisely as it happens in the classic analysis of
Collins and Kimball.

\section{The limit $R \to \infty$}
\label{sec:Rinf}

The distribution of the FPT remains well defined in the limit $R \to
\infty$ when the outer cylinder moves to infinity.  However, the MFPT
diverges in this limit.

We get $\alpha_n = \sqrt{\pi^2 n^2 + p L^2/D}$, and $g_n(r) \simeq
K_0(\alpha_n r/L) I_1(\alpha_n R/L)$, with an exponentially large
factor $I_1(\alpha_n R/L)$.  In particular, one finds
\begin{equation}
\tilde{S}_0(r;p) = \frac{1}{p}\left(1 - \frac{K_0(r\sqrt{p/D})}{K_0(\rho\sqrt{p/D})} \right)
\end{equation}
and
\begin{equation}
\frac{g_n(r)}{g'_n(\rho)} = - \frac{L}{\alpha_n} \, \frac{K_0(\alpha_n r/L)}{K_1(\alpha_n \rho/L)} \,.
\end{equation}
As a consequence, the Laplace-transformed pdf from
\eqref{eq:H_tilde} becomes
\begin{equation} \label{eq:t_solution_Rinf}
\tilde H(r,\theta; p) = \eta \, \frac{K_0(r\sqrt{p/D})}{K_0(\rho\sqrt{p/D})} 
+ 2 \eta \, L\sqrt{p/D} \,
\frac{K_1(\rho \sqrt{p/D})}{K_0(\rho \sqrt{p/D})} \sum\limits_{n=1}^\infty  \frac{K_0(\alpha_n r/L)}{\alpha_n K_1(\alpha_n \rho/L)}
\, \frac{\sin n\ve}{n\ve } \, \cos(n\theta) , 
\end{equation}
where
\begin{equation}
\eta^{-1} = 1 + \biggl(\frac{\pi D}{\kappa\, \ve} + \frac{L}{\pi} \Rve_\ve\biggr) \frac{\sqrt{p/D}\, K_1(\rho\sqrt{p/D})}{K_0(\rho\sqrt{p/D})}
\end{equation}
and
\begin{equation}
\Rve_\ve = 2\pi \sum\limits_{n=1}^\infty \frac{K_0(\alpha_n \rho/L)}{\alpha_n K_1(\alpha_n \rho/L)} \biggl(\frac{\sin n\ve}{n\ve}\biggr)^2 .
\end{equation}
In particular, the surface average of \eqref{eq:t_solution_Rinf}
yields
\begin{equation}   \label{eq:Hr_Rinf}
\overline{\tilde H(p)}_r = \eta \, \frac{K_0(r\sqrt{p/D})}{K_0(\rho\sqrt{p/D})}  \,,
\end{equation}
whereas the volume average in \eqref{lap_ex} diverges as $R\to
\infty$.

\subsection*{The particular case $\ve = \pi$}

The distribution of the FPT has already been studied in the particular
case of a perfectly absorbing inner cylinder (i.e., $\ve = \pi$ and
$\kappa = \infty$).  In this case, $\eta = 1$, and one retrieves the
Laplace-transformed pdf of the FPT to a cylinder of radius $\rho$
\cite{Spitzer58}.  In particular, the distribution of the FPT in this
particular case is known to be characterised by a power-law tail,
$H_\pi(r;t) \propto 1/(t \ln^2(t))$ as $t\to\infty$, where subscript
$\pi$ signifies that this solution corresponds to $\ve =
\pi$, in which the vertical coordinate $z$ (or $\theta$) is irrelevant
and thus the surface average does not change the solution.  Levitz
{\it et al.}
\cite{Levitz08} derived an approximation to $H_\pi(r;t)$ over the
whole range of times (see below)
\begin{equation}  \label{eq:H0_Levitz}
H_\pi(r; t) \simeq \begin{cases} \displaystyle
 \frac{r-\rho}{\sqrt{4\pi Dt^3}} \exp\bigl(- \frac{(r-\rho)^2}{4Dt} \bigr)  \hskip 20mm (t < \rho^2/(2D))  ,\cr
\displaystyle 
\frac{r/\rho-1}{2t \bigl(1 + \rho/\sqrt{2Dt}\bigr) \ln^2 \bigl((r + \sqrt{2Dt})/\rho\bigr)}  \quad (t > \rho^2/(2D)).  \end{cases}
\end{equation}
This approximation is only valid for $r$ close to $\rho$, i.e., when
the particle starts in a vicinity of the inner cylinder.

We consider the more general case of a partially reactive inner
cylinder (i.e., $\ve = \pi$ and $\kappa < \infty$), for which
\eqref{eq:t_solution_Rinf} becomes
\begin{equation}   \label{eq:Hr_Rinf1}
\tilde{H}_\pi(r;p) = \frac{K_0(r\sqrt{p/D})}{K_0(\rho\sqrt{p/D}) + \frac{D}{\kappa} \sqrt{p/D} K_1(\rho\sqrt{p/D})}  \,.
\end{equation}
Using the solution of an appropriate heat problem (see
Ref. \cite{Carslaw}, p. 337), we can write the inverse Laplace
transform as
\begin{equation}  \label{eq:Hpi}
H_\pi(r;t) = \frac{2\kappa}{\pi} \int\limits_0^\infty dq \, q \, e^{-Dt q^2}
\frac{Y_0(qr) (q J_1(q\rho) + hJ_0(q\rho)) - J_0(qr) (qY_1(q\rho) + h Y_0(q\rho))}{(qJ_1(q\rho)+hJ_0(q\rho))^2 + (qY_1(q\rho) + hY_0(q\rho))^2} \,,
\end{equation}
where $h = \kappa/D$.  In the long-time limit, the main contribution
to the integral comes from $q \approx 0$.  The asymptotic behaviour of
the integrand function at $q\to 0$ yields
\begin{equation}
H_\pi(r;t) \simeq \frac{2 \bigl(\frac{D}{\rho \kappa} + \ln(r/\rho)\bigr)}{t} 
 \int\limits_0^\infty \frac{dz\, e^{-z}}{\pi^2 + \bigl(-\frac{2D}{\rho\kappa} + \ln(z\rho^2/(4Dt)) + 2\gamma\bigr)^2} \,,
\end{equation}
where $\gamma \approx 0.5772\ldots$ is the Euler constant.  Discarding
a slowly varying function $\ln(z)$ in the denominator, we find the
following long-time asymptotic form
\begin{equation}  \label{eq:Hpi_long}
H_\pi(r;t) \simeq \frac{2\bigl(\frac{D}{\rho \kappa} + \ln(r/\rho)\bigr)}
{t \bigl[\pi^2 + \bigl(\ln(\rho^2/(4Dt)) + 2\gamma - \frac{2D}{\rho\kappa}\bigr)^2\bigr]} \,,
\end{equation}
which exhibits a very slow decrease $1/(t\ln^2(t))$ as $t\to\infty$.

In the short-time limit, one uses the asymptotic behaviour of the
integrand function as $q\to\infty$ to get
\begin{equation*}
H_\pi(r;t) \simeq \frac{2\kappa \sqrt{\rho/r}}{\pi} 
 \int\limits_0^\infty dq \, q \, e^{-Dtq^2} \frac{q \cos(q(r-\rho)) + h \sin(q(r-\rho))}{q^2 + h^2} \,. 
\end{equation*}
Ignoring the second term in the numerator, one gets the short-time
asymptotic behaviour
\begin{equation} \label{eq:Hpi_short}
H_\pi(r;t) \simeq \frac{\kappa\sqrt{\rho/r}}{\sqrt{\pi Dt}} \exp\biggl(-\frac{(r-\rho)^2}{4Dt}\biggr)  \quad (t\to 0).
\end{equation}

In the limit $\kappa\to\infty$ (perfect reactions), the exact solution
in \eqref{eq:Hpi} and its approximations \eqref{eq:Hpi_long} and
\eqref{eq:Hpi_short} become respectively
\begin{eqnarray}
H_\pi(r;t) &=& \frac{2D}{\pi} \int\limits_0^\infty dq \, q \, e^{-Dt q^2} \frac{Y_0(qr) J_0(q\rho) - J_0(qr) Y_0(q\rho)}{J_0^2(q\rho) + Y_0^2(q\rho)}  \,, \\
\label{eq:Hpi_asympt}
H_\pi(r;t) &\simeq& \frac{2\ln(r/\rho)}{t \bigl[\pi^2 + (\ln(\rho^2/(4Dt)) + 2\gamma)^2\bigr]}  \quad (t\to\infty)\,, \\
\label{eq:Hpi_asympt2}
H_\pi(r;t) &\simeq& \frac{(r-\rho)\sqrt{\rho/r}}{\sqrt{4\pi Dt^3}}\exp\biggl(-\frac{(r-\rho)^2}{4Dt}\biggr)  \quad (t\to 0).
\end{eqnarray}
As in Ref. \cite{Levitz08}, one can combine the short-time and
long-time approximations to cover the whole range of times.  If the
maximum of $H_\pi(r;t)$ occurs in the validity range of
\eqref{eq:Hpi_asympt2}, one can easily get the most probable FPT by
finding the zero of $\partial H_\pi(r;t)/\partial t$: $t_m =
(r-\rho)^2/(6D)$.  This value is three times smaller than that from
\eqref{eq:tm} for a partially reactive target.  The
difference in the prefactor comes from different power law corrections
to the common exponential function (cf. \eqref{eq:Hpi_short} and
\eqref{eq:Hpi_asympt2}): $t^{-3/2}$ for the perfectly reactive case
and $t^{-1/2}$ for the partially reactive case.

\begin{figure}
\includegraphics[width=85mm]{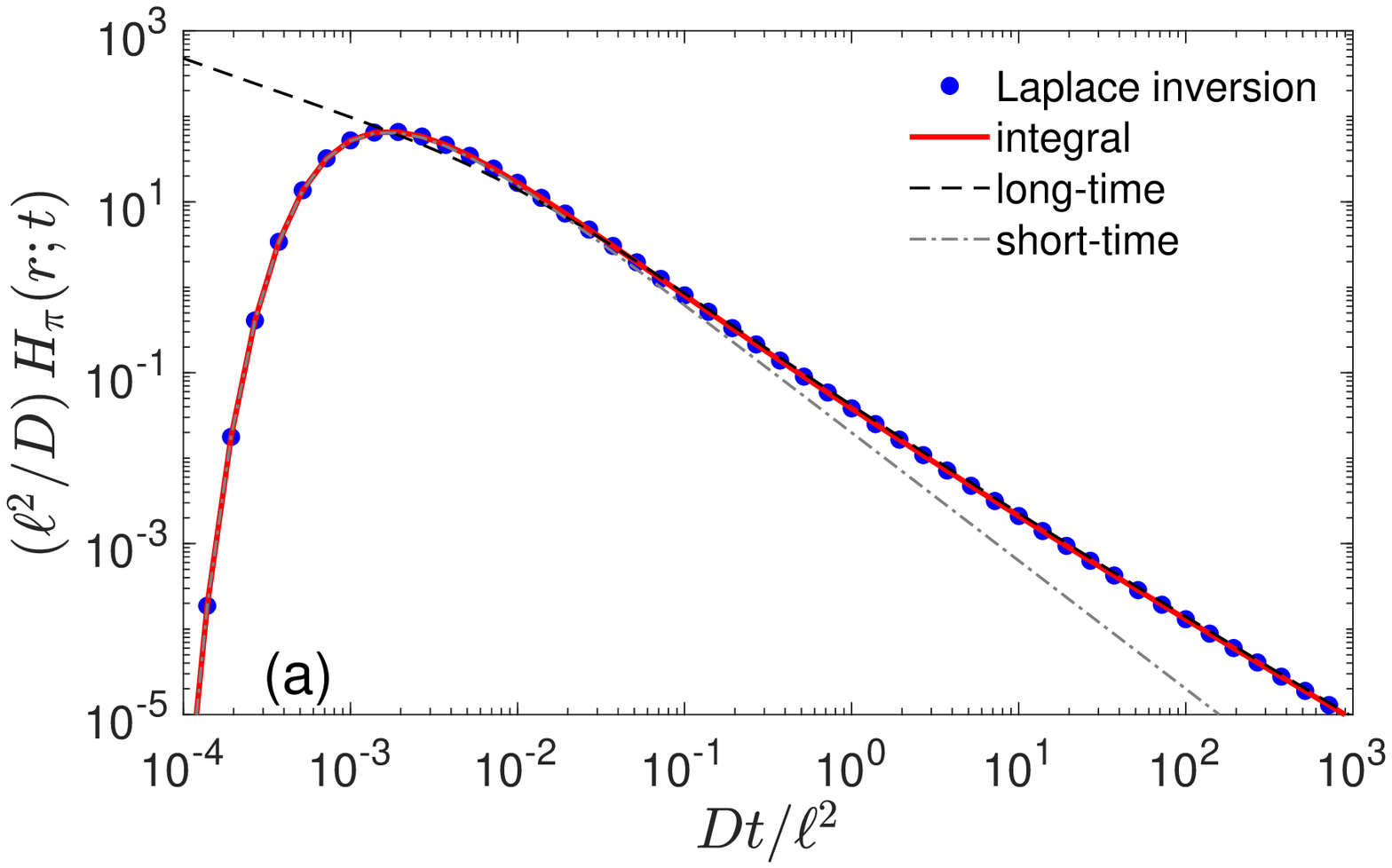}
\includegraphics[width=85mm]{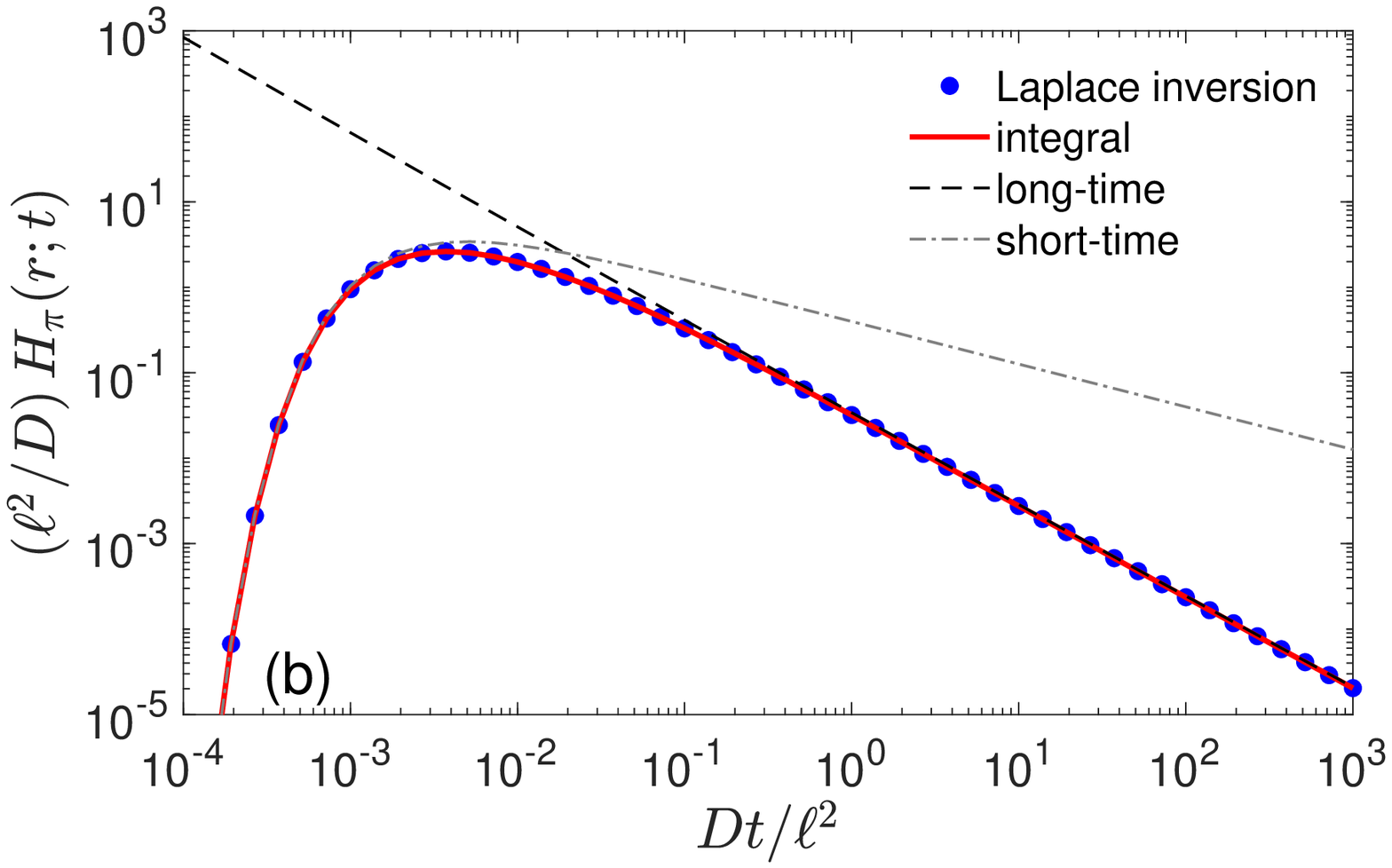}
\caption{The probability density $H_\pi(r;t)$ as function of $t$ for $R =
\infty$, $L/\ell = \ve = \pi$, $\rho/\ell = 0.1$, $\kappa = \infty$
{\bf (a)} and $\kappa \ell/D = 1$ {\bf (b)}. The exact integral
representation in \eqref{eq:Hpi} (solid line) is compared to the
numerical Laplace inversion of \eqref{eq:Hr_Rinf1} (circles) and to
the long-time and short-time approximations \eqref{eq:Hpi_long} and
\eqref{eq:Hpi_short}.
Length and time scales are fixed by setting $\ell = 1$ and
$\ell^2/D = 1$, with an auxiliary length $\ell = L/\pi$.}
\label{fig:Hpi}
% [Ht, t] = A_Oshanin3_Hr_t_Rinf_fig0;
\end{figure}

Figure \ref{fig:Hpi} illustrates the behaviour of the probability
density $H_\pi(r;t)$.  The exact integral representation in
\eqref{eq:Hpi} agrees perfectly with the numerical Laplace inversion
of \eqref{eq:Hr_Rinf1} over a very broad range of times, confirming
the high accuracy of the inversion procedure.  When the inner cylinder
is perfectly absorbing ($\kappa = \infty$), the long-time and
short-time approximations (\eqref{eq:Hpi_long} and
\eqref{eq:Hpi_short}) are very accurate and can be used to approximate
the probability density over the whole range of times.  For a
partially reactive cylinder ($\kappa = 1$), these approximations are
less accurate but still good.

\subsection*{Comment on the approximation in \eqref{eq:H0_Levitz}}
 
We briefly comment on the approximation in \eqref{eq:H0_Levitz}
derived in Ref. \cite{Levitz08}:

(i) we note that Eq. (1) in Ref. \cite{Levitz08} is incorrect: the
numerator and the denominator should be inter-changed.  This typing
error does not impact the consequent results.

(ii) the approximate Eq. (6) in Ref. \cite{Levitz08} was derived under
the simplifying assumption that the distance from the cylinder,
$\delta = r-\rho$, is much smaller than the radius of the cylinder
$\rho$.  The same derivation without this assumption yields
\begin{equation}  \label{eq:Levitz_extension}
H_\pi(r;t) = \frac{\ln(r/\rho)}{2t \bigl(1 + r/\sqrt{2Dt}\bigr) \ln^2 \bigl((r + \sqrt{2Dt})/\rho\bigr)} \,.
\end{equation}
When $\delta \ll \rho$, one has $\ln(r/\rho) = \ln(1 + \delta/\rho)
\simeq r/\rho - 1$, and thus retrieves \eqref{eq:H0_Levitz}, with
$r$ replaced by $\rho$ in the denominator.  As pointed out in Ref.
\cite{Levitz08}, the approximation $H_\pi(r;t)$ has the correct
normalisation by construction:
\begin{equation}
\int\limits_0^\infty dt \, H_\pi(r;t) = 1.
\end{equation}
Since this approximation is only valid for long times, it can be
completed by the short-time behaviour.

(iii) the approximation in \eqref{eq:H0_Levitz} and its extension in
\eqref{eq:Levitz_extension} are slightly different from our asymptotic
\eqref{eq:Hpi_asympt} but they become identical in the long-time limit.

\section{The limit $L\to\infty$}
\label{sec:Linf}

In the limit $L\to\infty$, the capped annular domain transforms into
an unbounded semi-infinite circular annulus: $\Omega = \{ \x \in \R^3
~:~ \rho < \sqrt{x^2 + y^2} < R, ~ z > 0\}$.  In this case, the
discrete summation variable $\xi = \pi n/L$ becomes continuous, and
one can replace sums by integrals.  In particular, we get $\alpha_n/L
= \sqrt{\xi^2 + p/D}$, and \eqref{eq:Rve} becomes
\begin{equation}  \label{eq:Rve_Linf}
\Rve_\ve = \int\limits_0^\infty d\xi \, \Upsilon\bigl(\rho, \sqrt{\xi^2+p/D}\bigr) \, \biggl(\frac{\sin \xi \epsilon}{\xi \epsilon}\biggr)^2 \,,
\end{equation}
with
\begin{equation}
\Upsilon(r, \alpha) 
= -2 \frac{I_0(\alpha r) K_1(\alpha R) + K_0(\alpha r) I_1(\alpha R)}{\alpha \bigl(I_1(\alpha \rho) K_1(\alpha R) - K_1(\alpha \rho) I_1(\alpha R)\bigr)} \,.
\end{equation}
Since \eqref{eq:eta} yields
\begin{equation}
\eta \simeq - \frac{\pi}{L} \frac{g_0(\rho)}{g'_0(\rho)} \biggl(\frac{\pi D}{\kappa \epsilon} + \Rve_\ve\biggr)^{-1}  ,
\end{equation}
the first term in \eqref{eq:H_tilde} vanishes, whereas the second term
transforms into
\begin{equation}  \label{eq:Htilde_Linf}
\tilde{H}(r,z;p) = \biggl(\frac{\pi D}{\kappa \epsilon} + \Rve_\ve\biggr)^{-1} \int\limits_0^\infty d\xi \, 
\Upsilon\bigl(r,\sqrt{\xi^2+p/D}\bigr) \frac{\sin \xi \epsilon}{\xi \epsilon} \cos \xi z .
\end{equation}
Due to the symmetry, this solution is also valid for an infinite
circular annulus with the target region $(-\epsilon,\epsilon)$ on the
inner cylinder.  Note also that in the second limit $R\to\infty$, the
function $\Upsilon$ reads
\begin{equation}
\Upsilon(r, \alpha) = \frac{2K_0(\alpha r)}{\alpha K_1(\alpha \rho)} \,.
\end{equation}

It is instructive to evaluate $\tilde{H}(r,z;p=0)$, which corresponds
to the normalisation of the FPT.  For a finite $R$, the integrals in
\eqref{eq:Rve_Linf} and \eqref{eq:Htilde_Linf} diverge as $p\to 0$,
implying $\tilde{H}(r,z;p=0) = 1$, as expected.  In contrast, when $R
= \infty$, both integrals are finite, and thus $\tilde{H}(r,z;p=0) <
1$, i.e., the density $H(r,z;t)$ is not normalised to $1$.  This is
the consequence of the transient character of the search process in
three dimensions (when $L = R = \infty$).  In this setting,
$\tilde{H}(r,z;p=0)$ is the probability of finding the target.

In the limit $p\to\infty$, we have
\begin{equation}
\Upsilon(r,\alpha) \simeq \frac{2\sqrt{\rho/r}}{\alpha} \, e^{-\alpha(r-\rho)}
\end{equation}
so that
\begin{equation}
\Rve_\ve \simeq \frac{\pi}{\epsilon\sqrt{p/D}} \,.
\end{equation}
For $0 < z < \epsilon$, we evaluate the leading contribution to the
integral in
\eqref{eq:Htilde_Linf} and get
\begin{equation}
\tilde{H}(r,z;p) \simeq \sqrt{\rho/r}\, \frac{e^{-(r-\rho)\sqrt{p/D}}}{1 + \sqrt{pD}/\kappa} \,,
\end{equation}
which does not depend on $\epsilon$ and $R$ in the leading order.
Inverting this relation, we get the short-time asymptotic behaviour
\begin{equation}   \label{eq:H_Linf_short}
H(r,z;t) \simeq \sqrt{\rho/r} \, e^{-(r-\rho)^2/(4Dt)} \, \biggl\{\frac{\kappa}{\sqrt{\pi Dt}} - \frac{\kappa^2}{D} 
\erfcx\biggl(\biggl(\frac{r-\rho}{\sqrt{4Dt}} + \frac{\kappa\sqrt{t}}{\sqrt{D}}\biggr)^2\biggr) \biggr\},
\end{equation}
where $\erfcx(x) = e^{x^2} \erfc(x)$ is the scaled complementary
error function.  In the perfectly reactive case $\kappa = \infty$,
this expression is reduced to
\begin{equation}  \label{eq:H_Linf_short0}
H(r,z;t) \simeq \sqrt{\rho/r} \, \frac{r-\rho}{\sqrt{4\pi Dt^3}}\,  e^{-(r-\rho)^2/(4Dt)} \,.
\end{equation}
In the case $z \geq \epsilon$, the integral in \eqref{eq:Htilde_Linf}
requires a more subtle evaluation that we do not discuss here.

In the opposite limit $p\to 0$, the major contribution to integrals
comes from $\xi \approx 0$.  Since
\begin{equation}
\Upsilon(r,\alpha) \simeq \frac{4\rho}{\alpha^2 (R^2 - \rho^2)} + \frac{2\rho}{R^2-\rho^2} Y_r  \qquad (\alpha \to 0),
\end{equation}
with
\begin{equation}
Y_r = \frac{r^2 - R^2}{2} - \frac{R^2+\rho^2}{4} + R^2 \ln(R/r) + \frac{R^2\rho^2 \ln(R/\rho)}{R^2-\rho^2} \,,
\end{equation}
we deduce the leading contribution to $\Rve_\ve$
\begin{equation}
\Rve_\ve \simeq \frac{2\pi \rho}{(R^2-\rho^2) \sqrt{p/D}} + \frac{\pi \rho Y_\rho}{\epsilon(R^2-\rho^2)} \,.
\end{equation}
Similarly, we get for $0 < z < \epsilon$
\begin{equation} 
\int\limits_0^\infty d\xi \, \Upsilon\bigl(r,\sqrt{\xi^2+p/D}\bigr) \frac{\sin \xi \epsilon}{\xi \epsilon} \cos \xi z \simeq 
 \frac{2\pi \rho}{(R^2-\rho^2) \sqrt{p/D}} + \frac{\pi \rho Y_r}{\epsilon(R^2-\rho^2)} \,,
\end{equation}
from which
\begin{equation}
\tilde{H}(r,z;p) \simeq 1 - \sqrt{p/D} \frac{1}{2\epsilon}
\biggl(\frac{D}{\kappa \rho} (R^2-\rho^2) + \frac{\rho^2 - r^2}{2} + R^2 \ln(r/\rho)\biggr),
\end{equation}
from which we deduce the long-time asymptotic behaviour
\begin{equation}  \label{eq:H_Linf_long}
H(r,z;t) \simeq \frac{1}{\sqrt{4\pi Dt^3}} \frac{1}{2\epsilon} 
\biggl(\frac{D}{\kappa \rho} (R^2-\rho^2) + \frac{\rho^2 - r^2}{2} + R^2 \ln(r/\rho)\biggr).
\end{equation}
We retrieved the characteristic $t^{-3/2}$ decay of the FPT density
for one-dimensional Brownian motion, which is supplemented by the
geometric information on the target and the annular domain.

\begin{figure}
\includegraphics[width=85mm]{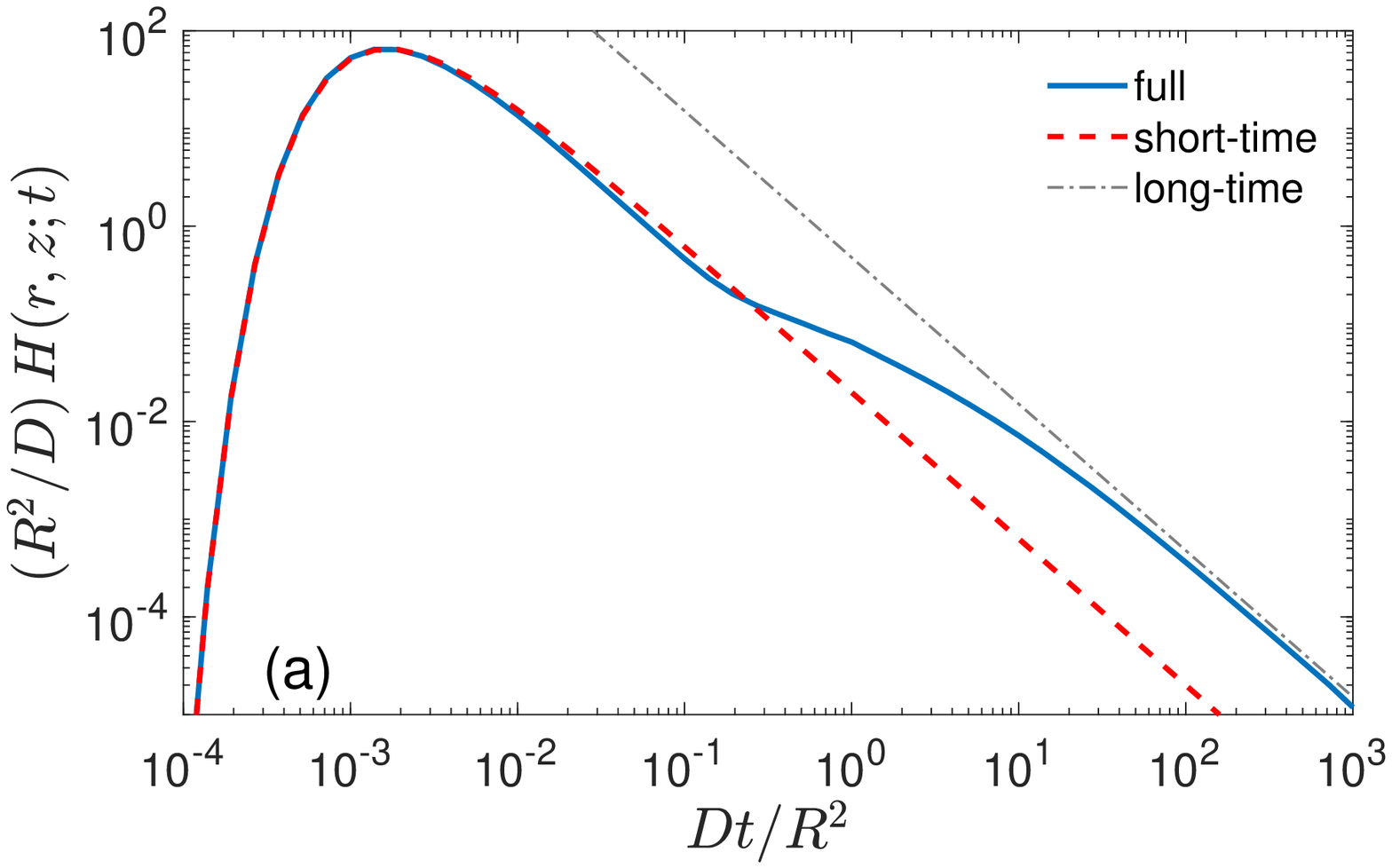}
\includegraphics[width=85mm]{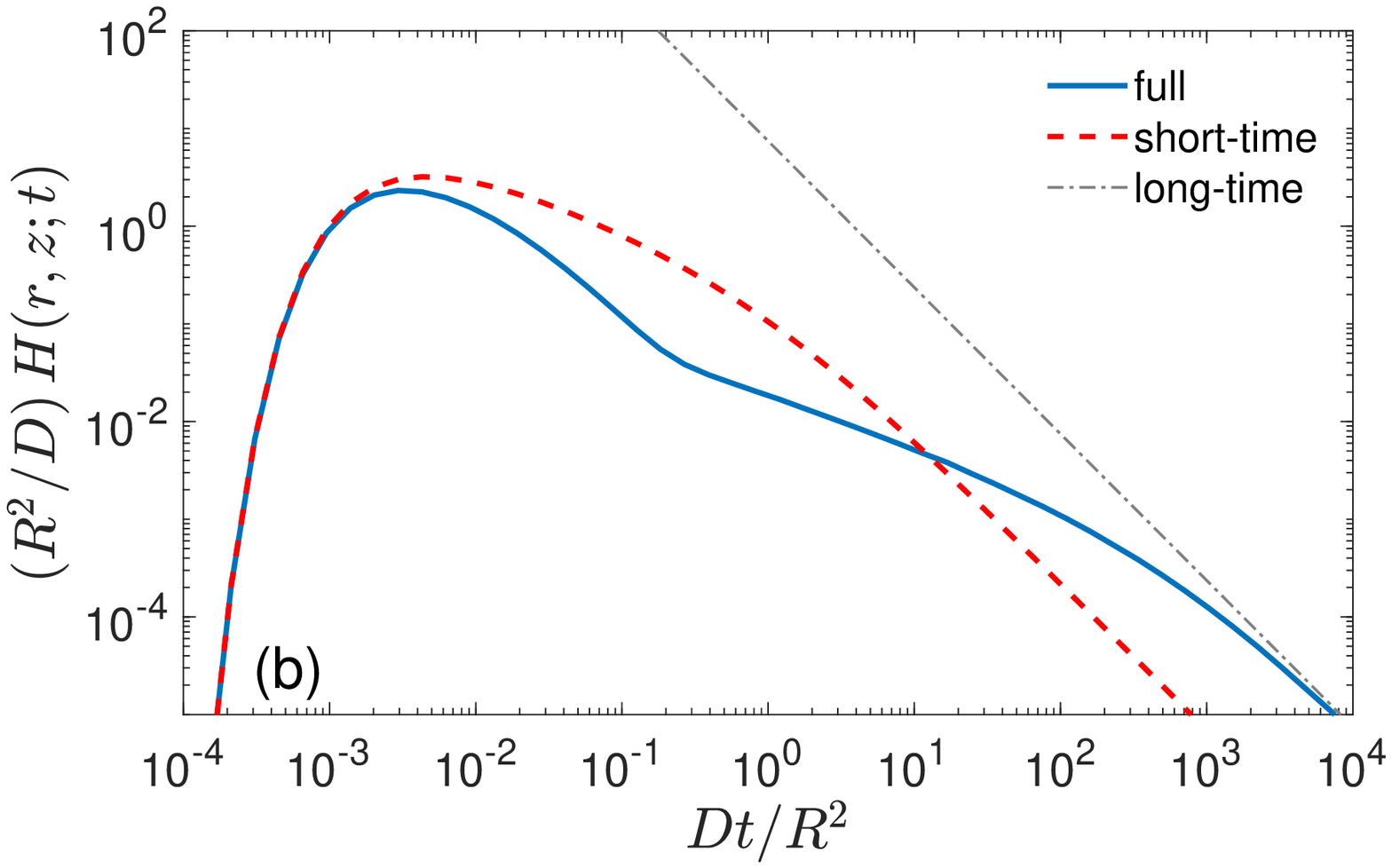}
\caption{The probability density $H(r,z; t)$ as function of $t$ for $L =
\infty$, $\epsilon/R = 0.2$, $z/R = 0.1$, $\rho/R = 0.1$, $r/R =
0.2$, $\kappa = \infty$ {\bf (a)} and $\kappa R/D = 1$ {\bf (b)}.
The numerical Laplace inversion of the integral representation in
\eqref{eq:Htilde_Linf} (solid line) is compared to its short-time
behaviour (dashed line) in \eqref{eq:H_Linf_short} (for $\kappa R/D =
1$) and \eqref{eq:H_Linf_short0} (for $\kappa = \infty$) and long-time
asymptotic relation in \eqref{eq:H_Linf_long} (dash-dotted line).
Length and time scales are fixed by setting $R = 1$ and $R^2/D=1$.}
\label{fig:H_t_Linf}
% [Ht, t] = A_Oshanin3_H_t_Linf_fig;
% [Ht, t] = A_Oshanin3_H_t_Linf_fig2;
\end{figure}

Figure \ref{fig:H_t_Linf} illustrates the probability density
$H(r,z;t)$ as function of $t$ for a semi-infinite circular annulus.
One can see that both short-time and long-time asymptotic relations
accurately capture this behaviour at small and large $t$, respectively.

\end{document}